\documentstyle[12pt,dina4p,epsfig]{article}
  \def\d{$\cdot$}
  \def\lp{\left. }
  \def\rp{\right. }
  \def\lr{\left( }
  \def\rr{\right) }
  \def\le{\left[ }
  \def\re{\right] }
  \def\lg{\left\{ }
  \def\rg{\right\} }

  \def\ts{\frac{-t}{s}}
  
  \def\us{\frac{-u}{s}}
  \def\tu{\frac{t}{u}}
  \def\ut{\frac{u}{t}}

  \newcommand{\beq}{\begin{equation}}
  \newcommand{\eeq}{\end{equation}}
  \newcommand{\bea}{\begin{eqnarray}}
  \newcommand{\eea}{\end{eqnarray}}
\title{
  \vskip-2cm
  {\baselineskip16pt
    \centerline{\normalsize \tt DESY 95-226 \hfill ISSN 0418-9833}
    \centerline{\normalsize \tt hep-ph/9511405 \hfill}
    \centerline{\normalsize \tt November 1995 \hfill}
  }
  \vskip2cm
  {\bf
    Inclusive Dijet Production at HERA: \\
    Direct Photon Cross Sections in Next-To-Leading Order QCD
  }
  \author{
    {M.\ Klasen,  G.\ Kramer} \\
    {II. Institut f\"ur Theoretische Physik}\thanks
    {Supported by Bundesministerium f\"ur Forschung und
     Technologie, Bonn, Germany under Contract 05\,6HH93P(5) and
     EEC Program "Human Capital and Mobility" through Network
     "Physics at High Energy Colliders" under Contract
     CHRX-CT93-0357 (DG12 COMA)} \\
     {Universit\"at Hamburg} \\
     {D - 22761 Hamburg, Germany}
    }
  \date{}
}
\begin{document}
\maketitle
\vspace{3cm}
\begin{abstract}
\thispagestyle{empty}
We have calculated inclusive two-jet cross sections in next-to-leading
order QCD for direct photoproduction in low $Q^2$ $ep$ collisions at HERA.
Infrared and collinear singularities in real and virtual contributions
are cancelled with the phase space slicing method. Analytical formulas
for the different contributions giving the dependence on the slicing
parameter are presented. Various one- and two-jet distributions have been
computed demonstrating the flexibility of the method.
\end{abstract}
\newpage
\setcounter{page}{1}
\section{Introduction}
Electron-proton scattering at HERA proceeds dominantly through
the exchange of photons with very small virtuality ($Q^2\simeq 0$).
An important fraction of the $\gamma p$ collisions contains high
transverse momentum ($E_T$) processes. The presence of this hard
momentum scale allows the application of perturbative QCD to
predict cross sections for the production of two or more high-$E_T$
jets. \\

In lowest order (LO) QCD, the hard scattering processes can be
classified in two types \cite{xxx1,xxx2}. In the so-called direct process,
the photon interacts in a point-like fashion with the quark which appears
either in the QCD Compton scattering $\gamma q\rightarrow gq$ or the
photon-gluon fusion process $\gamma g\rightarrow q\bar{q}$. In the
resolved process, the photon acts as a source of partons, which can
scatter with the partons coming out of the proton. These two processes
give two distinctly different types of high-$E_T$ event topologies.
The point-like interaction results in events with a characteristic
(2+1)-jet structure, i.e. two high-$E_T$ jets and one jet of low $E_T$
fragments of the proton. The resolved photon events have a characteristic
(2+2)-jet structure, since in addition to the two high-$E_T$ jets
and the low-$E_T$ proton remnant, fragments coming from the photon
will produce a second low-$E_T$ jet. While in the (2+1)-jet events the
total photon energy contributes to the hard scattering process, only a
fraction participates in the (2+2)-jet events. \\

Several methods for detecting the photon remnant jet have been proposed
and tested in Monte Carlo simulations, generally using a cut on the energy
in the photon direction or other kinematical variables \cite{xxx3,xxx4}.
Using an $E_T$ cluster algorithm, the ZEUS collaboration was actually able
to select events with two high-$E_T$ jets and a third cluster in the
approximate direction of the electron (photon) beam \cite{xxx5}.
They studied the properties of the photon remnant jet and found them to
agree with expectations. This shows that photoproduction events with
a photon remnant jet can be identified experimentally. \\

Unfortunately, this does not lead to a separation of the direct and
resolved processes, because their characterization described above is
only valid in LO QCD. In NLO, the direct cross section may also have
contributions with a photon remnant jet. Therefore its detection
does not uniquely separate the resolved cross section. In a complete
NLO calculation the direct contributions with a photon remnant jet
could be isolated, so that the remaining contributions could be compared
with the (2+1)-jet configuration. Of course, in this case the selection
procedure in the experimental analysis and in the theoretical calculation
must match. To make further progress in this direction, a complete
NLO calculation of the direct cross section is needed which is
flexible enough to incorporate the possible selection criteria for the
high-$E_T$ jets and the low-$E_T$ remnant jets. NLO calculations of direct
photoproduction have been done in the past for inclusive single hadron
\cite{xxx6}, inclusive single jet \cite{xxx7,xxx8,xxx9}, and for inclusive
dijet \cite{xxx7,xxx10} cross sections. Most of these results are applicable
only for special observables in inclusive single jet and dijet
production. \\

The leading order calculation of the direct component involves only
two subprocesses: $\gamma q \rightarrow gq$ and $\gamma g \rightarrow
q\bar{q}$. By including terms up to O($\alpha\alpha_s^2$), new contributions
arise. These are the one-loop contributions to $\gamma q \rightarrow gq$
and $\gamma g \rightarrow q\bar{q}$ and the three-body subprocesses
$\gamma q \rightarrow qgg$, $\gamma q \rightarrow qq\bar{q}$,
$\gamma q \rightarrow qq'\bar{q}'$, and $\gamma g \rightarrow gq\bar{q}$.
Besides the ultraviolet singularities in the one-loop contributions, which
can be removed by renormalization, additional infrared and collinear
singularities occur. They are extracted with the dimensional regularization
method. It is well known that for suitably defined inclusive observables
the infrared singularities in the one-loop contributions will cancel
against those in the three-body tree diagrams. The collinear singularities
from initial state radiation and $q\bar{q}$ production at the photon leg
can be factorized and absorbed in the respective parton distribution
functions. The hard collinear singularities from final state radiation
will cancel against the corresponding singularities in the one-loop
graphs. To achieve this technically and to maintain also the necessary
flexibility for new choices of observables, jet definitions, and
experimental cuts, we use the so-called phase space slicing method.
We introduce an invariant mass cut-off $y$, which allows us to
separate the phase space regions containing the singularities.
If $y$ is chosen sufficiently small, the integration over the
soft and collinear regions can be performed analytically in a
straightforward way. Then $y$ is just a technical cut which
needs not be connected with the experimental jet definition.
Of course, the suitably defined inclusive one- and two-jet
cross sections must be independent of $y$. Our method is very
similar to that of Baer et al. \cite{xxx7} who used two
distinct cut-off parameters for soft and collinear singularities.
The phase space slicing method with invariant mass cut-off has
been used for many processes. The calculation for photoproduction
proceeds along the same line as the calculation of higher order
QCD corrections for multijet cross sections in deep inelastic
$ep$ scattering first performed by Graudenz \cite{xxx11}. \\

The plan of the paper is as follows: In section 2 we review the
LO calculations and fix our notation. The techniques to perform
the NLO calculations are explained in section 3. In section 4 we
present some results for inclusive one- and two-jet cross sections
in order to demonstrate the flexibility of our method.
Section 5 contains a summary of our conclusions. In several
appendices we provide our analytically calculated results
needed for the Monte Carlo computations.
\section{Leading Order Cross Sections}
\subsection{Photon Spectrum}
The cross sections we have computed are essentially for kinematical
conditions as in the HERA experiments. There, electrons of $E_e = 26.7$ GeV
produce photons with small virtuality $Q^2=-q^2\simeq 0$, which then
collide with a proton beam of energy $E_p = 820$ GeV.
$q = p_e - p_e'$ is the momentum transfer of the electron to
the photon. The spectrum of the virtual photons
is approximated by the Weizs\"acker-Williams formula
\beq
 x_a F_{\gamma/e}(x_a) = \frac{\alpha}{2\pi}(1+(1-x_a)^2)\ln
                         \left(\frac{Q_{\max}^2 (1-x_a)}{m_e^2~x_a^2}\right),
\eeq
where $m_e$ is the electron mass and $x_a = \frac{pq}{pp_e} \simeq
E_\gamma / E_e$ is the fraction of the initial electron energy transferred
to the photon. $p$ is the four momentum of the incoming proton.
In the equivalent photon approximation, the cross section for the
process $ep \rightarrow e'X$ with arbitrary final state $X$ is then
given by the convolution
\beq
  \mbox{d}\sigma (ep \rightarrow e'X) = \int_{x_{a,\min}}^1 \mbox{d}x_a
  F_{\gamma/e}(x_a) \mbox{d}\sigma (\gamma p \rightarrow X),
\eeq
where d$\sigma (\gamma p \rightarrow X)$ denotes the cross section for
$\gamma p \rightarrow X$ with real photons of energy $E_\gamma = x_a E_e$.
\subsection{Jet Cross Sections in LO}
In leading order, the differential cross section for two-jet production
takes a very simple form. The partonic subprocesses are
$\gamma q \rightarrow gq$ and $\gamma g \rightarrow q\bar{q}$, which we
denote generically by $\gamma b \rightarrow p_1p_2$. $b$ is the parton
emitted from the proton with momentum $p_0 = x_b p$. Its distribution
function in the proton $F_{b/p}(x_b,M_b^2)$ depends on the
momentum fraction $x_b$ and
the factorization scale $M_b$. The final state partons have momenta
$p_1$ and $p_2$, which can be expressed by their transverse momentum
$E_T$ and their rapidities $\eta_1$ and $\eta_2$. The convention is
that the $z$ direction is parallel to the proton beam as used in the
experimental analysis. From energy and momentum conservation one
obtains
\bea
  x_a &=& \frac{E_T}{2E_e}\lr e^{-\eta_1} +e^{-\eta_2} \rr, \\
  x_b &=& \frac{E_T}{2E_p}\lr e^{ \eta_1} +e^{ \eta_2} \rr.
\eea
Thus, the kinematical variables of the two jets are related to the
scaling variables $x_a$ and $x_b$. For example, the rapidity $\eta_2$ of
the second jet is kinematically fixed by $E_T$, $\eta_1$, and the
photon fraction $x_a$ through the relation
\beq
  \eta_2 = -\ln\lr\frac{2x_aE_e}{E_T}-e^{-\eta_1}\rr.
\eeq
In the HERA experiments, $x_a$ is restricted to a fixed interval
$x_{a,\min} \leq x_a \leq x_{a,\max} < 1$. We shall disregard this
constraint and allow $x_a$ to vary in the kinematically allowed range
$x_{a,\min} \leq x_a \leq 1$, where
\beq
  x_{a,\min} = \frac{E_pE_Te^{-\eta_1}}{2E_eE_p-E_eE_Te^{\eta_1}}.
\eeq
{}From equations (3) and (4), we can express $x_b$ as a function of
$E_T$, $\eta_1$, and $x_a$:
\beq
  x_b = \frac{x_aE_eE_Te^{\eta_1}}{2x_aE_eE_p-E_pE_Te^{-\eta_1}}.
\eeq
This function is plotted in Fig.~1 for $E_T = 20$ GeV and four different
values of $\eta_1$ in the range $0 \leq \eta_1 \leq 3$ as a function of
$x_a$. Also shown is the envelope of these curves
\beq
  x_{b,\mbox{env}} = \frac{E_T^2}{x_aE_eE_p},
\eeq
which can be obtained from finding the minimum of $x_b$ with respect
to $\eta_1$ for fixed $x_a$.
The minimal $x_a$-values for $\eta_1 = 0$, 1, 2, 3 are 0.379, 0.143,
0.0557, and 0.0247.
The rapidity of the second jet is given by equation (5). \\

\begin{figure}[htbp]
 \begin{center}
  \begin{picture}(12,8)
   \epsfig{file=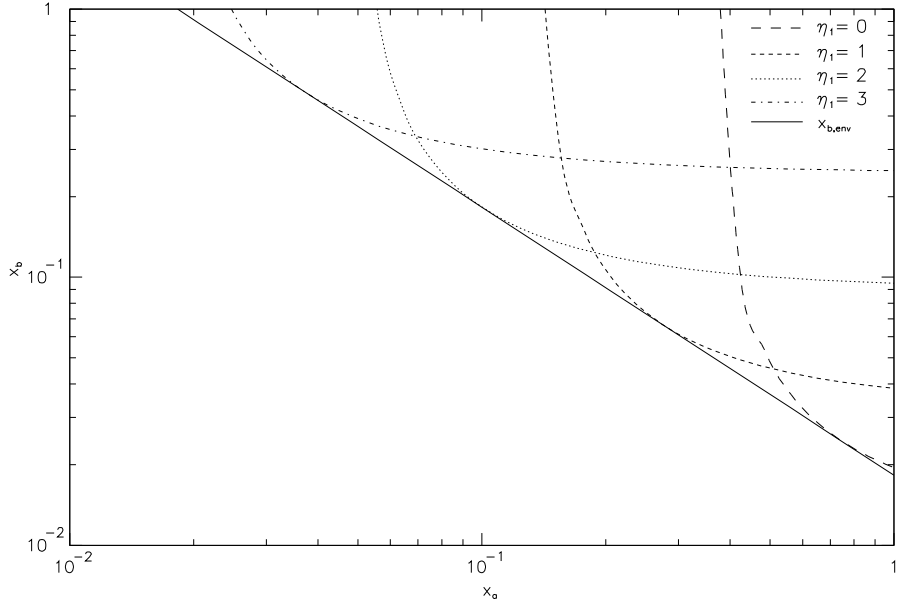,bbllx=105pt,bblly=95pt,bburx=520pt,bbury=710pt,%
           height=8cm,clip=,angle=-90}
  \end{picture}
  \caption{Contours of constant $\eta_1$ in the $x_a-x_b$ plane for
           $E_T = 20$ GeV. $x_{b,\mbox{env}}$ (full curve) is the envelope
           of the curves.}
 \end{center}
\end{figure}

Depending on $x_a$, different regions of $x_b$ are explored. Small values
of $x_b$ are reached for negative $\eta_1$ and the maximal possible $x_a$,
which is $x_a = 1$ in our case and $x_a = x_{a,\max}$ in the HERA
experiments. This is of interest for determining
the gluon structure function of the proton in the small $x$
region \cite{xxx12}.
Since under experimental conditions $x_a$ is not constrained very much,
it is possible to achieve minimal values of $x_b$ by selecting the
appropriate range in $E_T$, $\eta_1$, and $\eta_2$ according to (3)
and (4). \\

Of course, studying the gluon structure of the proton this way depends
very much on the cross sections in the appropriate regions.
For direct photoproduction, the two-jet cross section for
$ep \rightarrow e'+\mbox{jet}_1+\mbox{jet}_2+X$ is obtained from
\beq
  \frac{\mbox{d}^3\sigma}{\mbox{d}E_T^2\mbox{d}\eta_1\mbox{d}\eta_2}
  = \sum_b x_a F_{\gamma/e}(x_a) x_b F_{b/p}(x_b,M_b^2)
  \frac{\mbox{d}\sigma}{\mbox{d}t}(\gamma b \rightarrow p_1p_2).
\eeq
d$\sigma$/d$t$ stands for the differential cross section of the parton
subprocess $\gamma b \rightarrow p_1p_2$. The invariants of this process
are $s = (p_0+q)^2,~t = (p_0-p_1)^2$, and $u = (p_0-p_2)^2$.
They can be expressed by the final state variables $E_T$, $\eta_1$, and
$\eta_2$ and the initial state momentum fractions $x_a$ and $x_b$:
\bea
  s &=& 4 x_a x_b E_e E_p, \\
  t &=& -2 x_a E_e E_T e^{\eta_2} =  -2 x_b E_p E_T e^{-\eta_1}, \\
  u &=& -2 x_a E_e E_T e^{\eta_1} =  -2 x_b E_p E_T e^{-\eta_2}.
\eea
So, the dependence of the two-jet cross section on $E_T$, $\eta_1$,
and $\eta_2$ is determined through the
photon distribution function $F_{\gamma/e}$, the parton distribution
functions of the proton $F_{b/p}$, and the cross sections of the hard
subprocesses, which depend on $s$, $t$, and $u$. \\

For the inclusive
one-jet cross section, we must integrate over one of the rapidities
in (9). We integrate over $\eta_2$ and transform to the variable
$x_a$ using (3). The result is the cross section for
$ep \rightarrow e'+\mbox{jet}+X$, which depends on $E_T$ and $\eta$:
\beq
  \frac{\mbox{d}^2\sigma}{\mbox{d}E_T\mbox{d}\eta}
  = \sum_b \int_{x_{a,\min}}^1 \mbox{d}x_a x_a
  F_{\gamma/e}(x_a) x_b F_{b/p}(x_b,M_b^2)
  \frac{4E_eE_T}{2x_aE_e-E_Te^{-\eta}}
  \frac{\mbox{d}\sigma}{\mbox{d}t}(\gamma b \rightarrow p_1p_2).
\eeq
Here, $x_b$ is given by (7) with $\eta_1 = \eta$. \\

The cross sections for the two subprocesses
$\gamma q \rightarrow gq$ and $\gamma g \rightarrow q\bar{q}$ are
well known \cite{xxx13} and are given by
\bea
  \frac{\mbox{d}\sigma}{\mbox{d}t}(\gamma q_i \rightarrow gq_i) &=&
  \alpha\alpha_sQ_i^2 2C_F
  \frac{\pi}{s^2}
  \lr -\frac{u}{s}-\frac{s}{u}\rr, \\
  \frac{\mbox{d}\sigma}{\mbox{d}t}(\gamma g \rightarrow q_i\bar{q}_i) &=&
  \alpha\alpha_sQ_i^2
  \frac{\pi}{s^2}
  \lr \frac{u}{t}+\frac{t}{u}\rr,
\eea
where $C_F = (N_C^2-1)/(2N_C)$ and $N_C$ is the number of colors.
These cross sections are O($\alpha\alpha_s$). The index $i$ denotes
the quark flavor and $Q_i$ the quark charge. The magnitude of the
photon-gluon cross section determines whether one is sensitive to
the gluon structure function of the proton.
\section{Next-To-Leading Order Cross Sections}
The higher order corrections are calculated as usual
with the help of dimensional regularization. This method regularizes
the ultraviolet singularities in the one-loop contributions, which
are then subtracted using the modified minimal subtraction
($\overline{\mbox{MS}}$) scheme. Dimensional regularization can also
be used for the infrared and collinear singularities. In the following
subsections, we shall consider the virtual corrections and the various
real corrections needed to obtain finite cross sections in the
limit dimension $n \rightarrow 4$.
\subsection{Virtual Corrections up to O($\alpha\alpha_s^2$)}
The one-loop diagrams for $\gamma q \rightarrow gq$ and
$\gamma g \rightarrow q\bar{q}$ have an additional virtual
gluon, which leads to an extra factor $\alpha_s$. Interfering with
the LO diagrams of order O($\alpha\alpha_s$) produces the virtual
corrections to the $2 \rightarrow 2$ cross section up to
O($\alpha\alpha_s^2$). These contributions are well known for many years
from $e^+ e^- \rightarrow q\bar{q}g$ higher order QCD calculations
\cite{xxx14,xxx15}. For the corresponding photoproduction cross
section, one substitutes $Q^2 = 0$ and performs the necessary
crossings. The result can be found in \cite{xxx6,xxx16,xxx17}.
We have also compared with the results in \cite{xxx11} for deep
inelastic scattering $eq \rightarrow e'gq$ and
$e g \rightarrow e' q\bar{q}$, which can be expressed by
the invariants $s$, $t$, and $u$ after setting $Q^2 = 0$.
The structure of the corrections depends on
the color factors $C_F^2$, $C_F N_C$, and $C_F N_f$, where
$N_f$ is the number of flavors in the $q\bar{q}$ loops.
For completeness and for later use, we write the final result in the
form
\bea
  H_V(\gamma q_i \rightarrow g q_i) & = &
    e^2 g^2 \mu^{4\epsilon} 8(1-\epsilon) Q_i^2 \frac{\alpha_s}{2\pi}
    \left( \frac{4\pi\mu^2}{s} \right) ^\epsilon
    \frac{\Gamma(1-\epsilon)}{\Gamma(1-2\epsilon)}
    \frac{1}{4} \\
    && \left\{ C_F^2 V_{q1}(s,t,u) -\frac{1}{2} N_C C_F V_{q2}(s,t,u)
    \rp \nonumber \\ && \lp
    + C_F \lr \frac{1}{\epsilon} +\ln\frac{s}{\mu^2} \rr
    \left( \frac{1}{3} N_f - \frac{11}{6} N_C
    \right) T_q(s,t,u) \right\} + \mbox{O}(\epsilon) , \nonumber \\
  H_V(\gamma g \rightarrow q_i\bar{q}_i) & = &
    e^2 g^2 \mu^{4\epsilon} 8(1-\epsilon) Q_i^2 \frac{\alpha_s}{2\pi}
    \left( \frac{4\pi\mu^2}{s} \right) ^\epsilon
    \frac{\Gamma(1-\epsilon)}{\Gamma(1-2\epsilon)}
    \frac{1}{4} \\
    && \left\{ \frac{1}{2} C_F V_{g1}(s,t,u) -
    \frac{1}{4} N_C V_{g2}(s,t,u)
    \rp \nonumber \\ && \lp
    + \frac{1}{2}
    \lr \frac{1}{\epsilon} + \ln \frac{s}{\mu^2} \rr
    \left( \frac{1}{3} N_f - \frac{11}{6} N_C
    \right) T_g(s,t,u) \right\} + \mbox{O}(\epsilon)
. \nonumber \eea
$H_V$ gives the virtual corrections to the
corresponding reactions up to the $n$-dimensional phase space
factor
\beq
  \frac{\mbox{dPS}^{(2)}}{\mbox{d}t} =
  \frac{1}{\Gamma(1-\epsilon)}
  \left( \frac{4\pi s}{tu} \right) ^\epsilon
  \frac{1}{8\pi s}
\eeq
and the flux factor $1/(2s)$.
The expressions for $V_{q1}$, $V_{q2}$, $V_{g1}$, and $V_{g2}$ are
collected in appendix A. They contain singular terms
$\propto 1/\epsilon^2$ and $1/\epsilon$ ($2\epsilon = 4-n$),
which are always proportional to the LO cross sections
\bea
  T_q &=& (1-\epsilon)\left(
    - \frac{u}{s}-\frac{s}{u} \right)
    + 2\epsilon  , \label{g9} \\
  T_g &=&  (1-\epsilon) \left( \frac{t}{u}
    + \frac{u}{t} \right) -2\epsilon
\eea
(see eq. (14) and (15)). $g$ is the quark-gluon coupling and
$\mu$ is the renormalization scale. The terms proportional to
($\frac{1}{3} N_f - \frac{11}{6} N_C$) result from the renormalization
counter terms. The initial gluon spins are averaged by applying
a factor of $1/(2(1-\epsilon ))$.
\subsection{Real Corrections up to O($\alpha\alpha_s^2$)}
For the calculation of the hard scattering cross section up to
O($\alpha\alpha_s^2$), we must include all diagrams with an additional
parton in the final state, i.e. $\gamma q \rightarrow qgg$,
$\gamma q \rightarrow qq\bar{q}$, $\gamma q \rightarrow qq'\bar{q}'$,
and $\gamma g \rightarrow gq\bar{q}$. The four-vectors of these
subprocesses will be labeled by $qp_0\rightarrow p_1p_2p_3$,
where $q$ is the momentum of the incoming photon and $p_0$ is the
momentum of the incoming parton. The invariants will be denoted by
$s_{ij}=(p_i+p_j)^2$ and the previously defined Mandelstam variables
$s$, $t$, and $u$. For massless partons, the $2 \rightarrow 3$
contributions contain singularities at $s_{ij}=0$.
They can be extracted with the dimensional
regularization method and cancelled against those associated with
the one-loop contributions. \\

To achieve this we go through the following steps. First, the
matrix elements for the $2 \rightarrow 3$ subprocesses
are calculated in $n$ dimensions. The corresponding diagrams are
shown in Fig.~2 and are classified as I, II, ..., VII. They
are squared and averaged/summed over initial/final state spins
and colors and can be used for ingoing quarks, antiquarks, and gluons
with the help of crossing. The diagrams III' account for ghosts
to cancel unphysical polarizations of the gluons. This classification
is the same as in ref. \cite{xxx11}. Since the hadronic tensor is
symmetric, we need to consider only the products I$\cdot$I, II$\cdot$I,
III$\cdot$I etc. Furthermore, the products V$\cdot$IV, VI$\cdot$IV,
VII$\cdot$IV, VI$\cdot$V, VII$\cdot$V, and VII$\cdot$VI are not
singular and
vanish with the cut-off describing the boundary of the singular
region. Such terms will be neglected consistently in the following. \\

\begin{figure}[htbp]
 \begin{center}
  \begin{picture}(12,8)
   \epsfig{file=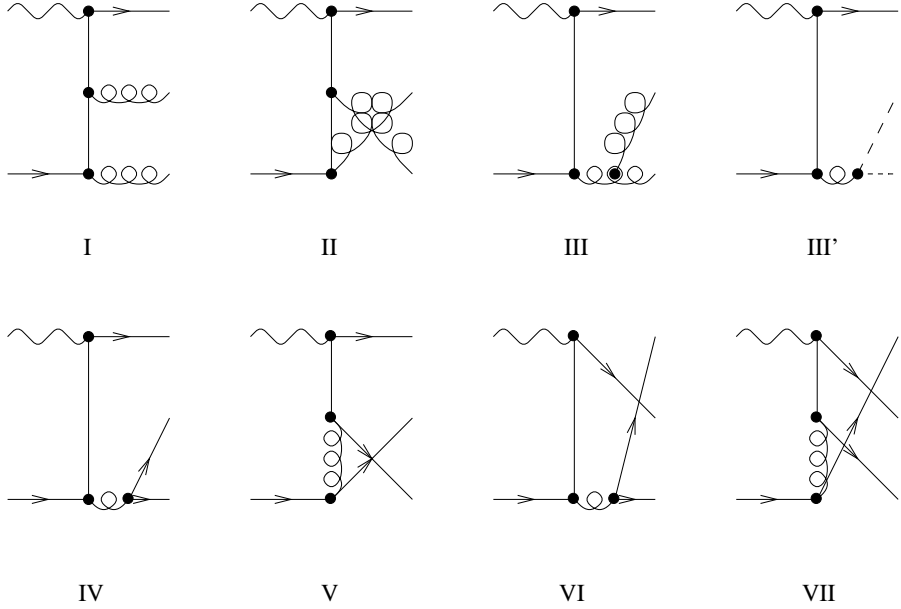,bbllx=38pt,bblly=0pt,bburx=572pt,bbury=795pt,%
           height=8cm,clip=,angle=-90}
  \end{picture}
  \caption{Three-body diagrams for classes I, ..., VII as described in
           the text.}
 \end{center}
\end{figure}

Next, one can distinguish three classes of singularities in
photoproduction ($Q^2 \simeq 0$) depending on which propagators
in the squared matrix elements vanish. Examples for these three
classes are shown in Fig.~3. The X marks the propagator leading
to the divergence. In the first graph, it is the invariant $s_{23}$
given by momenta of the final state. Therefore, this singularity
will be called final state singularity. The second graph becomes
singular for $s_{q1}=(q+p_1)^2=0$, when the photon and the final
quark momentum are parallel. This is the class of photon initial
state singularities. In the third graph, the singularity occurs at
$s_{03}=(p_0+p_3)^2$, where $p_0$ is the initial parton momentum.
This stands for parton initial state singularity.
The first class is familiar from similar calculations for jet
production in $e^+e^-$ annihilation \cite{xxx16}, the third class
from jet production in deep inelastic $ep$ scattering ($Q^2\neq 0$)
\cite{xxx11}. The second class occurs only for photoproduction
\cite{xxx9,xxx10}. \\

\begin{figure}[t]
 \begin{center}
  \begin{picture}(10,4)
   \epsfig{file=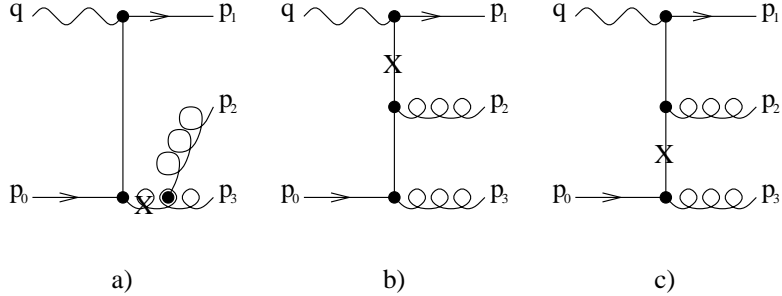,bbllx=190pt,bblly=85pt,bburx=430pt,bbury=705pt,%
           height=4cm,clip=,angle=-90}
  \end{picture}
  \caption{Three-body diagrams with final and initial state singularities.}
 \end{center}
\end{figure}

When squaring the sum of all diagrams in Fig.~2, we encounter
terms, where more than one of the invariants become singular, e.g.
when one of the gluon momenta $p_3 \rightarrow 0$ so that
$s_{23}=0$ and $s_{03}=0$. These infrared singularities are
disentangled by a partial fractioning decomposition, so that
every term has only one vanishing denominator. This also allows
the separation of the different classes of singularities in Fig.~3.
Except for additional terms separating the photon initial state
singularity, we follow the procedure in \cite{xxx11}. It turns out
that the results are always proportional to the LO cross sections
involved in the hard scatter like $T_q$ and $T_g$ in eq. (19) and (20)
\beq
  H_{F,I,J} = K T_i.
\eeq
Here, $F, I$, and $J$ denote final state, photon initial state and
proton initial state contributions. \\

As the last step, the decomposed matrix elements have to be integrated
up to $s_{ij} \leq y s$. $y$ characterizes the region, where the
two partons $i$ and $j$ cannot be resolved.
Then, the singular kernel $K$ produces terms
$\propto 1/\epsilon^2$ and $1/\epsilon$, which will cancel against
those in the virtual diagrams or be absorbed into structure functions,
and finite corrections proportional to $\ln^2 y$, $\ln y$, and $y^0$.
Terms of O($y$) will be neglected. In the following, we shall give
the results for the different classes of singularities separately.
\subsubsection{Final State Singularities}
In this section, we assume that after partial fractioning
the $2 \rightarrow 3$ matrix elements are singular only for $s_{12}=0$.
For the integration over the singular region of phase space, we
choose as the coordinate system the c.m. system of partons $p_1$ and
$p_2$. The angles of the other parton three-momenta $p_0$ and $p_3$
with respect to $p_1$ and $p_2$ are shown in Fig.~4. $\chi$ is
the angle between $p_0$ and $p_3$, $\theta$ is the angle between
$p_0$ and $p_1$, and $\phi$ is the azimuthal angle between the planes
defined by $p_0$ and $p_1$ and $p_0$ and $p_3$, respectively. \\

\begin{figure}[htbp]
 \begin{center}
  \begin{picture}(4.5,5)
   \epsfig{file=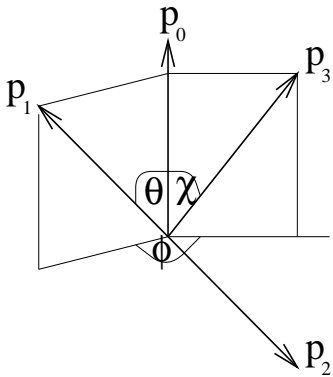,bbllx=205pt,bblly=303pt,bburx=413pt,bbury=488pt,%
           height=5cm,clip=,angle=-90}
  \end{picture}
  \caption{Diagram for the three-body final state defining the kinematical
           variables in the c.m. system of partons $p_1$ and $p_2$.}
 \end{center}
\end{figure}

In general, the invariants
\bea
  s &=& (p_0+q)^2, \\
  t &=& (p_0-p_1-p_2)^2-2p_1p_2, \\
  u &=& (p_0-p_3)^2
\eea
differ from the corresponding two-body invariants (see eq. (10)-(12)),
but approach them for $p_2=0$ or $p_2$ parallel to $p_1$. Then,
$t=(p_0-\bar{p}_1)^2$ and $u=(p_0-p_3)^2$, where $\bar{p}_1=p_1+p_2$ is
the four-momentum of the recombined jet and $p_3$ is the four-momentum
of the second jet. This produces a jet mass
$\bar{p}_1^2=(p_1+p_2)^2\neq 0$, which vanishes in the limit
$y \rightarrow 0$. For $p_2=0$, we have $\bar{p}_1=p_1$, and for
$p_2$ parallel to $p_1$ we have $\bar{p}_1 \propto p_1$. The
definitions of $t$ and $u$ are unique only in the limit
$s_{12}\rightarrow 0$. Other choices for the three-parton final
state variables differ by finite contributions of O($y$), which are
neglected throughout. We substitute the angle $\theta$ by
\beq
  b = \frac{1}{2}(1-\cos\theta)
\eeq
and introduce the new variable
\beq
  z' = \frac{p_1p_2}{p_0q}.
\eeq

The three-body phase space in $n$ dimensions can be factorized into
two two-body phase space terms \cite{xxx11}
\beq
  \mbox{dPS}^{(3)} = \mbox{dPS}^{(2)} \mbox{dPS}^{(r)},
\eeq
where
\beq
  \frac{\mbox{dPS}^{(2)}}{\mbox{d}t}
  = \frac{1}{\Gamma(1-\epsilon)}
  \left( \frac{4\pi s}{tu} \right) ^\epsilon
  \frac{1}{8\pi s}
\eeq
and
\beq
  \mbox{dPS}^{(r)} =
  \left( \frac{4\pi}{s} \right) ^\epsilon \frac{\Gamma (1-\epsilon)}
  {\Gamma (1-2\epsilon)} \frac{s}{16 \pi ^2} \frac{1}{1-2\epsilon}
  \mbox{d}\mu_F
\eeq
with
\beq
  \mbox{d}\mu_F =
  \mbox{d}z' z'^{-\epsilon} \left( 1+\frac{z's}{t} \right)
  ^{-\epsilon}
  \frac{\mbox{d}b}{N_b} b^{-\epsilon} (1-b)^{-\epsilon}
  \frac{\mbox{d}\phi}{N_\phi} \sin^{-2\epsilon} \phi.
\eeq
$N_b$ and $N_\phi$ are normalization factors:
\bea
  N_b & = & \int_0^1 \mbox{d}b (b(1-b))^{-\epsilon}
  = \frac{\Gamma ^2 (1-\epsilon)}{\Gamma (2-2\epsilon)}, \\
  N_\phi & = &\int_0^\pi \sin^{-2\epsilon} \phi \mbox{d}\phi
  = \frac{\pi 4^\epsilon \Gamma (1-2\epsilon)}{\Gamma ^2 (1-\epsilon)}.
\eea
$\mbox{dPS}^{(2)}$ is the same phase space factor as in the two-body
LO final state. The full range of integration in
$\mbox{dPS}^{(r)}$ is given by $z' \in [0,-t/s]$, $b \in [0,1]$, and
$\phi \in [0,\pi ]$. The singular region is defined by the
requirement that partons $p_1$ and $p_2$ are recombined, i.e.
$p_2 = 0$ or $p_2$ parallel to $p_1$, so that $s_{12}=0$.
We integrate over this region up to $s_{12} \leq y s$, which restricts
the range of $z'$ to $0 \leq z' \leq \min\{ -t/s, y \} \equiv y_F$. \\

The matrix elements, which are obtained by squaring the sum of
diagrams in Fig.~2, are expressed by the variables $s$, $t$, $u$,
$z'$, $b$, and $\phi$.
Depending on the structure of the interfering diagrams and the color
factors, we distinguish seven types of contributions: $F_1$, $F_2$, ...,
$F_7$. Five of them are for an incoming quark or antiquark and
two for an incoming gluon. The definition of the $F_i$ with the
corresponding color factors can be read off Tab. 1. \\

\begin{table}[htbp]
\begin{center}
\begin{tabular}{|c|c|c|c|}
\hline
  Class & Inc. Parton & Diagrams & Color Factor \\
\hline
  $F_1$  & Quark   & I\d I, II\d II, II\d I
    & $N_C C_F^2 / N_C $ \\
  $F_2$  & Quark   & II\d I, III\d I, III\d II
    & $-1/2\: N_C^2 C_F / N_C $ \\
  $F_3$  & Quark   & III\d I, III\d II
    & $-1/2\: N_C^2 C_F / N_C $ \\
  $F_4$  & Quark   & III\d III
    & $N_C^2 C_F / N_C $ \\
  $F_5$  & Quark   & IV\d IV, V\d V, VI\d VI, VII\d VII
    & $1/2\: N_C C_F / N_C $ \\
  $F_6$  & Gluon   & I\d I, II\d II, II\d I
    & $N_C C_F^2 / 2 N_C C_F $ \\
  $F_7$  & Gluon   & II\d I, III\d I, III\d II
    & $-1/2\: N_C^2 C_F / 2 N_C C_F $ \\
\hline
\end{tabular}
\end{center}
\caption{Classification of the $2\rightarrow 3$ matrix elements for
         final state singularities}
\end{table}

In class $F_1$, we have QED-type diagrams proportional to $C_F^2$, where
a gluon can become soft or collinear with a quark. $F_2$ contains
QED-type diagrams and interference terms with diagrams containing
a three-gluon vertex proportional to $C_FN_C$. Class $F_3$ stands for
contributions of this latter type, where the infrared divergence is
caused by the two-gluon final state. $F_4$ has only contributions
from non-abelian diagrams, and $F_5$ contains the four-quark diagrams.
In this class, the divergence originates from the collinear
$q\bar{q}$ pair. In $F_6$ and $F_7$, we have contributions with
incoming gluons proportional to $C_F$ and $N_C$, respectively.
We emphasize that this classification rests upon the use of the
Feynman gauge for the gluon spin summation with all ghost contributions
included. Of course, the sum of all terms is gauge invariant. \\

After integrating the $2 \rightarrow 3$ matrix elements with
final state singularities $H_{F_i}$, we obtain
\beq
  \int \mbox{dPS}^{(r)} H_{F_i} =
    e^2 g^2 \mu^{4\epsilon} 8(1-\epsilon) Q_i^2
    \frac{\alpha_s}{2\pi}
    \left( \frac{4\pi\mu^2}{s} \right) ^\epsilon
    \frac{\Gamma(1-\epsilon)}{\Gamma(1-2\epsilon)}
    \frac{1}{4} F_i + \mbox{O}(\epsilon).
\eeq

The final results for the $F_i$ are written down in appendix B.
They contain the infrared and collinear singularities which cancel
against those in the virtual corrections. At this point, it is essential
that the singular terms are all proportional to the LO matrix
elements $T_q(s,t,u)$ and $T_g(s,t,u)$ and that the variables
$s$, $t$, and $u$ defined in (22)-(24) correspond to the two-body
variables in the infrared and collinear limit with an appropriate
definition of the recombined jet momentum.
\subsubsection{Photon Initial State Singularities}
To integrate over the singularity $s_{q3} = 0$, we use the same
$p_1$, $p_2$ c.m. coordinate system as in the last section.
Let us assume that we have isolated the terms in the three-body cross
sections with a singularity at $s_{q3} = 0$, i.e. where the outgoing
quark momentum $p_3$ is collinear with the photon momentum $q$.
In this case, the quark $p_3$ is part of the photon remnant and $p_1$
and $p_2$ are the parton momenta of the two-body subprocess.
We introduce a new variable
\beq
  z_a = \frac{p_1p_2}{p_0q} \in [X_a,1],
\eeq
where $X_a=\frac{p_1p_2}{p_0p_e}\simeq E_q/E_e$ is now the fraction
of the initial electron energy transferred to the quark in the hard
scatter. Then, $z_a$ parametrizes the phase space of the collinear
quark $p_3=(1-z_a)q$ and renders the Mandelstam variables in the
form
\bea
  s &=& (p_0+z_aq)^2, \\
  t &=& (p_0-p_1)^2, \\
  u &=& (p_0-p_2)^2.
\eea
With the additional variable
\beq
  z'' = \frac{qp_3}{qp_0},
\eeq
the three-body phase space factorizes into
\beq
  \mbox{dPS}^{(3)} = \mbox{dPS}^{(2)} \mbox{dPS}^{(r)}.
\eeq
$\mbox{dPS}^{(2)}$ is identical to (28), and
\beq
  \mbox{dPS}^{(r)} =
  \left( \frac{4\pi}{s}\right) ^\epsilon \frac{\Gamma (1-\epsilon)}
  {\Gamma (1-2\epsilon)} \frac{s}{16 \pi ^2} H_a(z'')
  \mbox{d}\mu_I,
\eeq
where
\beq
  \mbox{d}\mu_I =
  \mbox{d}z'' z''^{-\epsilon}
  \frac{\mbox{d}z_a}{z_a} \lr \frac{z_a}{1-z_a}\rr ^\epsilon
  \frac{\mbox{d}\phi}{N_\phi} \sin^{-2\epsilon}\phi
  \frac{\Gamma(1-2\epsilon)}{\Gamma^2(1-\epsilon)}
\eeq
and
\beq
  H_a(z'') = \lr 1+\frac{z''}{z_a}\rr ^{-1+2\epsilon}
  \lr 1- \frac{z''}{1-z_a} \rr ^{-\epsilon}
  = 1+ \mbox{O}(z'')
\eeq
can be approximated by 1 as it leads only to negligible terms of O($y$).
The full region of integration is given by
$z'' \in [0,-u/s]$, $z_a \in [X_a,1]$, and
$\phi \in [0,\pi ]$, whereas the singular region is defined by the
requirement that parton $p_3$ is collinear to $q$, so that $s_{q3} < y s/z_a$.
This restricts the range of $z''$ to $0 \leq z'' \leq \min\{ -u/s, y \}
\equiv y_I$. \\

The matrix elements from Fig.~2 can be put into seven classes $I_1$, $I_2$,
..., $I_7$ displayed in Tab. 2. The classification is again according to
incoming parton, color factor, and general structure.

\begin{table}[htbp]
\begin{center}
\begin{tabular}{|c|c|c|c|c|}
\hline
  Class & Inc. Parton & Diagrams & Color Factor \\
\hline
  $I_1$  & Quark   & I\d I, II\d II, II\d I
    & $N_C C_F^2 / N_C $ \\
  $I_2$  & Quark   & II\d I, III\d I, III\d II
    & $-1/2\: N_C^2 C_F / N_C $ \\
  $I_3$  & Quark   & IV\d IV, V\d V, VI\d VI, VII\d VII
    & $1/2\: N_C C_F / N_C $ \\
  $I_4$  & Quark   & IV\d IV, V\d V, VI\d VI, VII\d VII
    & $1/2\: N_C C_F / N_C $ \\
  $I_5$  & Quark   & IV\d IV, V\d V, VI\d VI, VII\d VII
    & $C_F / N_C $ \\
  $I_6$  & Gluon   & I\d I, II\d II, II\d I
    & $N_C C_F^2 / 2 N_C C_F $ \\
  $I_7$  & Gluon   & II\d I, III\d I, III\d II
    & $-1/2\: N_C^2 C_F / 2 N_C C_F $ \\
\hline
\end{tabular}
\end{center}
\caption{Classification of the $2 \rightarrow 3$ matrix elements for
         photon initial state singularities}
\end{table}

In all classes, the incoming photon can become collinear to a quark line
only, since there is no direct coupling to a gluon line. Furthermore,
there are only single poles, i.e. only collinear singularities. The
three-body matrix elements are factorized into LO two-body parton-parton
scattering matrix elements and a singular kernel $K$. Since the
quark (antiquark) produced by the photon becomes part of the photon
remnant, the remaining two-body process is induced by an antiquark (quark)
in the initial state, which interacts with a quark or antiquark of the
same or different flavor or a gluon from the proton. The factorized
LO parton-parton scattering cross sections are given by the
functions $U_i(s,t,u)$ $(i=1,2,3)$ related to the various processes as in
Tab. 3. \\

\begin{table}[htbp]
\begin{center}
$
\begin{array}{|c|l|}
\hline
  \mbox{Process} & \mbox{Matrix element}
  \: |{\cal M}|^2 = 8 N_C C_F g^4 U_i \\
\hline
  q\overline{q} \rightarrow gg             &  U_1(s,t,u) \\
  qq'           \rightarrow qq'            &  U_2(s,t,u) \\
  q\overline{q}'\rightarrow q\overline{q}' &  U_2(u,t,s) \\
  q\overline{q} \rightarrow q'\overline{q}'&  U_2(t,s,u) \\
  qq            \rightarrow qq             &
    U_2(s,t,u)+U_2(s,u,t)+U_3(s,t,u) \\
  q\overline{q} \rightarrow q\overline{q}  &
    U_2(u,t,s)+U_2(u,s,t)+U_3(u,t,s) \\
  qg            \rightarrow qg             & -U_1(t,s,u) \\
\hline
\end{array}
$
\end{center}
\caption{LO parton-parton scattering matrix elements}
\end{table}

The functions $U_i(s,t,u)$ are decomposed with respect to their color
structure as follows:
\bea
  U_1 & = & \left( C_F-N_C\frac{ut}{s^2} \right)
  (1-\epsilon)
  \left[ (1-\epsilon) \left( \frac{t}{u} + \frac{u}{t} \right)
  -2\epsilon \right] \\
  &=&  C_F T_{1a}(s,t,u)
       - \frac{N_C}{2} T_{1b}(s,t,u)  , \nonumber \\
  U_2 & = & \frac{1}{2} \left( \frac{s^2+u^2}{t^2}
  -\epsilon \right) = \frac{1}{2} T_2(s,t,u)  , \\
  U_3 & = & -\frac{1}{N_C} (1-\epsilon)
  \left( \frac{s^2}{tu} + \epsilon \right) = \frac{1}{N_C} T_3(s,t,u)
. \eea

The functions $T_{1a}, T_{1b}, T_2$, and $T_3$ appear in the
final result
\beq
  \int \mbox{dPS}^{(r)}
  H_{I_i} =
  \int_{X_a}^1 \frac{\mbox{d}z_a}{z_a}
  e^2 g^2 \mu^{4\epsilon} 8(1-\epsilon)\frac{\alpha_s}{2\pi}
  \left( \frac{4\pi\mu^2}{s} \right) ^\epsilon
  \frac{\Gamma(1-\epsilon)}{\Gamma(1-2\epsilon)}
  \frac{1}{4} I_i +\mbox{O}(\epsilon)
, \eeq
which is obtained from the integration of the decomposed
matrix elements $H_{I_i}$ over $\mbox{dPS}^{(r)}$.
It is clear that the photon initial state singularities lead to
contributions familiar from LO resolved photoproduction. \\

The terms $I_i$ are collected in appendix C and show explicitly
the poles in $1/ \epsilon$ proportional to
\beq
  P_{q_i \leftarrow \gamma}(z_a) =
  2N_C Q_i^2 \frac{1}{2} \left[ z_a^2+(1-z_a)^2 \right].
\eeq
This function appears in the evolution equation of the photon structure
function as an inhomogeneous or so-called point-like term. Therefore,
the photon initial state singularities
can be absorbed into the photon structure function. The necessary
steps are well known \cite{xxx16,xxx18}.
We define the renormalized distribution
function of a parton $a$ in the electron $F_{a/e}(X_a,M_a^2)$ as
\beq
  F_{a/e} (X_a,M_a^2) =
  \int_{X_a}^1 \frac{\mbox{d}z_a}{z_a}
  \left[ \delta_{a\gamma } \delta (1-z_a) + \frac{\alpha}{2\pi}
  R_{a\leftarrow \gamma }(z_a, M_a^2)\right] F_{\gamma /e}
  \lr\frac{X_a}{z_a}\rr
, \eeq
where $R$ has the general form
\beq
  R_{a \leftarrow \gamma } (z_a, M_a^2) =
 -\frac{1}{\epsilon}P_{a\leftarrow \gamma }(z_a)\frac{\Gamma (1-\epsilon)}
  {\Gamma (1-2\epsilon)} \left( \frac{4\pi\mu^2}{M_a^2} \right)
  ^\epsilon + C_{a\leftarrow \gamma} (z_a)
\eeq
with $C = 0$ in the $\overline{\mbox{MS}}$ scheme.
The renormalized partonic cross section for $\gamma b \rightarrow \mbox{jets}$
is then calculated from
\beq
  \mbox{d}\sigma(\gamma b\rightarrow \mbox{jets}) = \mbox{d}\bar{\sigma}
  (\gamma b\rightarrow \mbox{jets}) - \frac{\alpha}{2\pi} \int
  \mbox{d}z_a R_{a\leftarrow \gamma }(z_a,M_a^2)
  \mbox{d}\sigma (ab\rightarrow \mbox{jets})
. \eeq
d$\bar{\sigma} (\gamma b\rightarrow \mbox{jets})$ is the higher order
cross section before the subtraction procedure, and
d$\sigma (ab\rightarrow \mbox{jets})$ contains the LO parton-parton
scattering matrix elements $U_i(s,t,u)$.
The factor $\frac{4\pi\mu^2}{M_a^2}$ in (49) is combined with the factor
$\frac{4\pi\mu^2}{s}$ in (46) and leads to $M_a^2$ dependent terms
of the form
\beq
  -\frac{1}{\epsilon}P_{a\leftarrow \gamma }(z_a)\le
  \lr\frac{4\pi\mu^2}{s}\rr^\epsilon
  -\lr\frac{4\pi\mu^2}{M_a^2}\rr^\epsilon\re
  = -P_{a\leftarrow \gamma }(z_a) \ln\lr\frac{M_a^2}{s}\rr,
\eeq
so that the cross section after subtraction in (50) will depend on the
factorization scale $M_a^2$.
\subsubsection{Parton Initial State Singularities}
The computation of the parton initial state contributions is very similar
to the last section. We again use the $p_1$, $p_2$ c.m. coordinate system
as defined in Fig.~4. Singularities occur in this case for $s_{03}=0$,
when the incoming parton $p_0$ radiates a soft gluon ($p_3=0$) or
a collinear parton $p_3$.
The new variable
\beq
  z_b = \frac{p_1p_2}{p_0q} \in [X_b,1]
\eeq
describes the momentum of parton $p_3=(1-z_b)p_0$ and the reduced
momentum of the parton involved in the hard scatter $z_bp_0$.
$X_b=\frac{p_1p_2}{Pq} \simeq E_{q/g}/E_p$ is the energy fraction of the
incoming proton involved in the $2 \rightarrow 2$ subprocess. The
corresponding invariants are
\bea
  s &=& (z_bp_0+q)^2, \\
  t &=& (z_bp_0-p_1)^2, \\
  u &=& (z_bp_0-p_2)^2.
\eea
Using the variable
\beq
  z'''=\frac{p_0p_3}{p_0q},
\eeq
we can again factorize the three-body phase space into
\beq
  \mbox{dPS}^{(3)} = \mbox{dPS}^{(2)} \mbox{dPS}^{(r)},
\eeq
where $\mbox{dPS}^{(2)}$ stands for (28).
\beq
  \mbox{dPS}^{(r)} =
  \left( \frac{4\pi}{s}\right) ^\epsilon \frac{\Gamma (1-\epsilon)}
  {\Gamma (1-2\epsilon)} \frac{s}{16 \pi ^2} H_b(z''')
  \mbox{d}\mu_J,
\eeq
where
\beq
  \mbox{d}\mu_J =
  \mbox{d}z''' z'''^{-\epsilon}
  \frac{\mbox{d}z_b}{z_b} \lr \frac{z_b}{1-z_b}\rr ^\epsilon
  \frac{\mbox{d}\phi}{N_\phi} \sin^{-2\epsilon}\phi
  \frac{\Gamma(1-2\epsilon)}{\Gamma^2(1-\epsilon)}
\eeq
and
\beq
  H_b(z''') = \lr 1-z'''\rr ^{-1+2\epsilon}
  \lr 1- \frac{z'''}{1-z_b} \rr ^{-\epsilon}
  = 1+ \mbox{O}(z''')
\eeq
can again be approximated by 1. The integration over
$z''' \in [0,-u/s]$, $z_b \in [X_b,1]$, and
$\phi \in [0,\pi ]$ is restricted to the singular region of
$z'''$ in the range $0 \leq z''' \leq \min\{ -u/s, y \}
\equiv y_J$. \\

The three-body matrix elements can be computed from Fig.~2 and classified
according to color factor, incoming parton, and type of parton participating
in the two-body process as in Tab. 4. In the case of final state
singularities considered in section 3.2.1, the ingoing parton (quark or gluon)
was always the ingoing parton of the two-body subprocess. As could already
been seen in the last section, this is not the case for initial state
singularities. For example, if an ingoing quark is collinear to the
outgoing gluon (quark), the factorized 2-jet matrix element has an
initial quark (gluon). \\

\begin{table}[htbp]
\begin{center}
\begin{tabular}{|c|c|c|c|c|}
\hline
  Class & Inc. Parton & Inc. Parton & Diagrams & Color Factor \\
         & (3-jet) & (2-jet) &               &                    \\
\hline
  $J_1$  & Quark   & Quark   & I\d I, II\d II, II\d I
    & $N_C C_F^2 / N_C $ \\
  $J_2$  & Quark   & Quark   & II\d I, III\d I, III\d II
    & $-1/2\: N_C^2 C_F / N_C $ \\
  $J_3$  & Quark   & Gluon   & IV\d IV, V\d V, VI\d VI, VII\d VII
    & $1/2\: N_C C_F / N_C $ \\
  $J_4$  & Gluon   & Quark   & I\d I, II\d II, II\d I
    & $N_C C_F^2 / 2 N_C C_F $ \\
  $J_5$  & Gluon   & Quark   & II\d I, III\d I, III\d II
    & $-1/2\: N_C^2 C_F / 2 N_C C_F $ \\
  $J_6$  & Gluon   & Gluon   & II\d I, III\d I, III\d II
    & $-1/2\: N_C^2 C_F / 2 N_C C_F $ \\
  $J_7$  & Gluon   & Gluon   & III\d III
    & $N_C^2 C_F / 2 N_C C_F $ \\
\hline
\end{tabular}
\end{center}
\caption{Classification of the $2 \rightarrow 3$ matrix elements for parton
         initial state singularities}
\end{table}

In Tab. 4, $J_1$ is the class of QED type terms, where a gluon in the
final state is collinear with the initial state quark. The same occurs
in $J_2$, where QED-type and non-abelian diagrams interfere. $J_3$
contains the four-quark diagrams, in which an outgoing quark is collinear
with an ingoing quark. In $J_4$ and $J_5$, we have an ingoing gluon
and an outgoing collinear quark. The classes differ only by their color
factors. In $J_6$ and $J_7$, an outgoing gluon is collinear with the
ingoing gluon, where $J_6$ contains QED-type and interference terms and
$J_7$ contains pure non-abelian terms. \\

The next step is to perform the $z'''$ integration with the upper limit
$y_J$. The result can be written in the form
\beq
  \int \mbox{dPS}^{(r)}
  H_{J_i} =
  \int_{X_b}^1 \frac{\mbox{d}z_b}{z_b}
  e^2 g^2 \mu^{4\epsilon} 8(1-\epsilon)Q_i^2\frac{\alpha_s}{2\pi}
  \left( \frac{4\pi\mu^2}{s} \right) ^\epsilon
  \frac{\Gamma(1-\epsilon)}{\Gamma(1-2\epsilon)}
  \frac{1}{4} J_i +\mbox{O}(\epsilon)
, \eeq
where the $J_i$ are given in appendix D. Besides
$s$, $t$, $u$, and $y_J$, they still depend on the integration variable
$z_b$. The singular parts take the form
\bea
  J_{1}&=& \le -\frac{1}{\epsilon}\frac{1}{C_F} P_{q\leftarrow q}(z_b)
            +\delta (1-z_b) \lr \frac{1}{\epsilon^2}
            +\frac{1}{\epsilon} \lr -\ln \ts +\frac{3}{2} \rr \rr
            \re C_F^2T_q(s,t,u) +\mbox{O}(\epsilon^0), \\
  J_{2}&=& \le \delta (1-z_b) \frac{1}{\epsilon} \ln \ut
           \re \lr-\frac{1}{2}N_CC_F\rr T_q(s,t,u) +\mbox{O}(\epsilon^0), \\
  J_{3}&=& \le-\frac{1}{\epsilon}\frac{1}{C_F}P_{g\leftarrow q}(z_b)
           \re \frac{C_F}{2}T_g(s,t,u) +\mbox{O}(\epsilon^0), \\
  J_{4}&=& \le -\frac{2}{\epsilon} P_{q\leftarrow g}(z_b)
            \re C_FT_q(s,t,u) +\mbox{O}(\epsilon^0), \\
  J_{6}&=& \le \frac{2}{\epsilon}\frac{1}{N_C}P_{g\leftarrow g}(z_b)
            +\delta(1-z_b)
            \lr -\frac{2}{\epsilon^2}+\frac{1}{\epsilon}
            \lr \ln\frac{tu}{s^2}-\frac{2}{N_C}
            \lr\frac{11}{6}N_C-\frac{1}{3}N_f\rr\rr\rr\re \\
       &&   \lr -\frac{N_C}{4}\rr T_g(s,t,u) +\mbox{O}(\epsilon^0).
            \nonumber
\eea
$J_5$ is O($\epsilon$) and does not contribute. $J_6$ contains
contributions from $J_6$ as well as $J_7$.
Some of the $J_i$ contain infrared singularities $\propto 1
/\epsilon^2$, which must cancel against the corresponding singularities
in the virtual contributions. The singular parts are decomposed in such a
way that the Altarelli-Parisi kernels in four dimensions proportional to
$1/\epsilon$ are split off. They also appear in the evolution equations for
the parton distribution functions and are defined explicitly in appendix D.
The singular terms proportional to these kernels are absorbed as usual
into the scale dependent structure functions
\beq
  F_{b/p} (X_b,M_b^2)  =
  \int_{X_b}^1 \frac{\mbox{d}z_b}{z_b}
  \left[ \delta_{bb'} \delta (1-z_b) + \frac{\alpha_s}{2\pi}
  R'_{b\leftarrow b'} (z_b, M_b^2) \right] F_{b'/p}
  \lr\frac{X_b}{z_b}\rr
, \eeq
where $F_{b'/p}\lr\frac{X_b}{z_b}\rr$ is the LO structure function before
absorption of the collinear singularities and
\beq
  R'_{b \leftarrow b'} (z_b, M_b^2) =
  -\frac{1}{\epsilon} P_{b\leftarrow b'} (z_b) \frac{\Gamma (1-\epsilon)}
  {\Gamma (1-2\epsilon)} \left( \frac{4\pi\mu^2}{M_b^2} \right)
  ^\epsilon + C'_{b\leftarrow b'} (z_b)
\eeq
with $C' = 0$ in the $\overline{\mbox{MS}}$ scheme. Then, the renormalized
higher order hard scattering cross section d$\sigma (\gamma b \rightarrow
$jets) is calculated from
\beq
  \mbox{d}\sigma(\gamma b\rightarrow \mbox{jets}) = \mbox{d}\bar{\sigma}
  (\gamma b\rightarrow \mbox{jets}) - \frac{\alpha_s}{2\pi} \int
  \mbox{d}z_b R'_{b\leftarrow b'}(z_b,M_b^2)
  \mbox{d}\sigma (\gamma b'\rightarrow \mbox{jets})
. \eeq
d$\bar{\sigma} (\gamma b\rightarrow \mbox{jets})$ is the higher order
cross section before the subtraction procedure, and
d$\sigma (\gamma b'\rightarrow \mbox{jets})$ contains the lowest order
matrix elements $T_q(s,t,u)$ and $T_g(s,t,u)$ in $n$ dimensions. This well
known factorization prescription \cite{xxx18} removes finally all remaining
collinear singularities. It is universal and leads for all processes
to the same definition of structure functions if the choice concerning the
regular function $C'$ in (68) is kept fixed. Similar to the case of photon
initial state singularities, the higher order cross sections in (69)
will depend on the factorization scale $M_b$ due to terms of the form
$P_{b\leftarrow b'}(z_b) \ln \lr \frac{M_b^2}{s} \rr$.
\subsection{Jet Cross Sections in NLO}
To obtain a finite cross section, we must add the four parts considered
in sections 3.1 and 3.2.1-3.2.3. Then, the poles in $1/\epsilon$ must
cancel, and we can take the limit $\epsilon \rightarrow 0$. The result
is a special kind of two-jet cross section, where the recombination
of two partons into one jet or the recombination of a parton with the
photon or proton remnant jet is done with an invariant mass cut-off $y$.
In LO, this cross section was written down in (9). Including the NLO
corrections, it has the form
\bea
  \frac{\mbox{d}^3\sigma}{\mbox{d}E_T^2\mbox{d}\eta_1\mbox{d}\eta_2}&=&
  \sum_b x_a F_{\gamma /e}(x_a)
  \le x_bF_{q/p}(x_b, M_b^2) \frac{\mbox{d}\sigma}{\mbox{d}t}
  (\gamma q\rightarrow p_1p_2)
  + \frac{\mbox{d}\tilde{\sigma}}{\mbox{d}t}
  (\gamma q\rightarrow p_1p_2) \rp \\
  &&\hspace*{2.4cm}+ x_bF_{g/p}(x_b, M_b^2) \frac{\mbox{d}\sigma}{\mbox{d}t}
  (\gamma g\rightarrow p_1p_2)
  + \frac{\mbox{d}\tilde{\sigma}}{\mbox{d}t}
  (\gamma g\rightarrow p_1p_2) \nonumber \\
  &&\hspace*{2.36cm}\lp + x_bF_{b/p}(x_b,M_b^2)
  \frac{\mbox{d}\tilde{\sigma}}{\mbox{d}t}
  (ab \rightarrow p_1p_2) \re
  . \nonumber
\eea
In (70), $\frac{\mbox{d}\textstyle\sigma}{\mbox{d}\textstyle t}
(\gamma q\rightarrow p_1p_2)$ and
$\frac{\mbox{d}\textstyle\sigma}{\mbox{d}\textstyle t}
(\gamma g\rightarrow p_1p_2)$ stand for
two-body contributions in LO and NLO together with analytically integrated
contributions of the soft and collinear divergent regions of the three-parton
final state. The contributions from the photon initial state
singularities are denoted $\frac{\mbox{d}\textstyle \tilde{\sigma}}
{\mbox{d}\textstyle t}
(ab \rightarrow p_1p_2)$.
$\frac{\mbox{d}\textstyle \tilde{\sigma}}{\mbox{d}\textstyle t}
(\gamma q\rightarrow p_1p_2)$ and
$\frac{\mbox{d}\textstyle \tilde{\sigma}}{\mbox{d}\textstyle t}
(\gamma g\rightarrow p_1p_2)$
originate from the integration of the soft and collinear singularities
of the parton initial state. \\

The quark induced cross section can be written as
\beq
  \frac{\mbox{d}\sigma}{\mbox{d}t}
  (\gamma q\rightarrow p_1p_2)
    = \alpha_s(\mu^2)CT_q+\frac{\alpha_s^2(\mu^2)}{2\pi}
    C(T_qA_q+B_q)
, \eeq
where
\bea
  A_q & = &
    -\frac{1}{6}\lr 11N_C-2N_f \rr \ln\frac{s}{\mu^2}
    +\frac{N_f}{3}\lr\ln y_F-\frac{5}{3}\rr
    -N_C\frac{11}{6}\ln y_F \\ &&
    +\lr 2C_F-N_C \rr \le \frac{1}{2}\ln^2\tu-\frac{1}{4}\ln^2\ts
    +\ln y_F\ln\ts-\mbox{Li}_2\lr\frac{y_F s}{t}\rr \re \nonumber \\ &&
    {}+N_C\le -\frac{1}{2}\ln^2 y_F-\frac{1}{2}\ln^2 \frac{y_F s}
    {-u}+\frac{1}{4}\ln^2\us-\mbox{Li}_2\lr \frac{y_F s}{u}\rr
    +\frac{67}{18}-\frac{\pi^2}{2} \re \nonumber \\ &&
    {}+C_F\le \frac{\pi^2}{3}-\frac{7}{2}-\ln^2 y_F-\frac{3}{2}
    \ln y_F \re , \nonumber \\
  B_q & = &
    -3C_F\frac{s}{u}\ln\us-\frac{1}{2}\lr 2C_F-N_C \rr
    \le 2 \ln\frac{-us}{t^2}+\lr 2+\frac{u}{s}\rr\lr
    \pi^2+\ln^2\frac{t}{u}\rr \rp \\ &&
    \lp {}+ \lr 2+\frac{s}{u}\rr\ln^2\ts \re
    , \nonumber \\
  T_q & = &
    -  \frac{u}{s}-\frac{s}{u} , \mbox{~and~}
    C = \frac{2\pi\alpha}{s^2} Q_i^2 C_F . \eea
For the gluon induced cross section, we have
\bea
  \frac{\mbox{d}\sigma}{\mbox{d}t}
  (\gamma g\rightarrow p_1p_2)
    &=&\frac{3}{8}\alpha_s(\mu^2)CT_g+\frac{3}{8}
    \frac{\alpha_s^2(\mu^2)}{2\pi}
    C(T_gA_g+B_g)
, \eea
where
\bea
  A_g & = &
    -\frac{1}{6}\lr 11N_C-2N_f \rr \ln\frac{s}{\mu^2}+C_F
    \le \ln^2\ts+\ln^2\us-2\ln^2 y_F-3\ln y_F\re \\ &&
    +N_C\le -\frac{1}{2}\ln^2\frac{tu}{s^2}-\frac{\pi^2}{6}
    -\frac{1}{2}\ln^2\frac{y_F s}{-t} -\frac{1}{2}\ln^2
    \frac{y_F s}{-u} +\ln^2y_F
    +\frac{1}{4}\ln^2\ts+\frac{1}{4}\ln^2\us\rp \nonumber \\
    && {}\lp
    -\mbox{Li}_2\lr\frac{y_F s}{t}\rr-\mbox{Li}_2\lr\frac{y_Fs}{u}\rr \re
    , \nonumber \\
  B_g & = &
    3 C_F \lr \frac{t}{u}\ln\us+\frac{u}{t}\ln\ts\rr
    +\frac{1}{2}\lr 2C_F-N_C\rr
    \le 2\ln\frac{tu}{s^2}+\lr 2+\frac{u}{t}\rr\ln^2\us
    \rp \\ && {} \lp
    +\lr 2+\frac{t}{u}\rr\ln^2\ts\re
    , \nonumber \\
  T_g & = &
     \frac{t}{u} + \frac{u}{t}
. \eea
The cut dependence of the final state NLO corrections in $A_q$ and $A_g$ is
contained in the $y_F$ dependent terms. For $y_F \rightarrow 0$, these terms
behave like $(-\ln^2 y_F)$, which leads to unphysical negative cross sections
for very small $y_F$.
Thus if $y$ is used as a physical cut, it must be sufficiently large. In
most of the applications, we shall use these results for computing inclusive
cross sections, in which the $y$ dependence of the two-jet cross section
cancels against the $y$ dependence of the numerically calculated three-jet
cross section. The quadratic terms in $(\ln y_F)$ are connected with the
soft divergence of the three-parton final state. \\

The contributions from the photon initial state singularities necessarily
have a photon remnant. They depend on the cut-off $y_I$ only through
terms proportional to $(\ln y_I)$. This contribution can be considered
separately only if $y_I$ is a physical cut, i.e. if it is sufficiently
large. In case one introduces other kinematical conditions to define
the photon remnant terms in the NLO direct cross section, one must add
the appropriate three-body contributions. The two-jet cross section is
calculated from
\bea
  \frac{\mbox{d}\tilde{\sigma}}{\mbox{d}t}(ab \rightarrow p_1p_2)
  & = &
  \frac{\mbox{d}\hat{\sigma}}{\mbox{d}t}(ab \rightarrow p_1p_2)
  \frac{\alpha}{2\pi} \\
  &&
  \int_{X_a}^1 \frac{\mbox{d}z_a}{z_a} \le
  P_{q_i\leftarrow\gamma}(z_a)\lr\ln\lr\frac{(1-z_a)y_Is}{z_aM_a^2}\rr-1\rr+
  2N_C\frac{Q_i^2}{2}\re \nonumber
. \eea
In (79), $M_a$ is the factorization scale at the photon leg. This dependence
must cancel against the $M_a$ dependence of the LO resolved contribution
(see eq. (48)). This has been demonstrated explicitly in \cite{xxx19} for
the inclusive single jet cross section. The
$\frac{\mbox{d}\textstyle \hat{\sigma}}{\mbox{d}\textstyle t}
(ab\rightarrow p_1p_2)$
are the cross sections for the $2 \rightarrow 2$ subprocesses
which were given in section 3.2.2 in terms of the functions $U_i(s,t,u)$.
These are the well known LO cross sections also used for the calculation
of hadron-hadron cross sections. \\

{}From the parton initial state singularities, we deduce
\bea
  \frac{\mbox{d}\tilde{\sigma}}{\mbox{d}t}
  (\gamma q\rightarrow p_1p_2)
  & = &
    \frac{\alpha_s^2(\mu^2)}{2\pi}CT_q \\
  && \lg
    \int_{X_b}^1 \frac{\mbox{d}z_b}{z_b}X_b
    F_{q/p}\lr \frac{X_b}{z_b},M_b^2\rr
    \le C_F \tilde{J}_1(z_b) + P_{q\leftarrow q}(z_b)\ln\frac{s}{M_b^2}
    \re \rp \nonumber \\
  && \hspace*{-0.16cm} \lp
    +\int_{X_b}^1\frac{\mbox{d}z_b}{z_b}X_b
    F_{g/p}\lr\frac{X_b}{z_b},M_b^2\rr\le
    P_{q\leftarrow g}(z_b)\lr\ln\lr\frac{(1-z_b)y_Js}{z_bM_b^2}\rr-1\rr+
    \frac{1}{2}\re \rg \nonumber , \\
  \frac{\mbox{d}\tilde{\sigma}}{\mbox{d}t}
  (\gamma g\rightarrow p_1p_2)
  & = &
    \frac{3}{8}\frac{\alpha_s^2(\mu^2)}{2\pi }CT_g \\
  && \lg
    \int_{X_b}^1 \frac{\mbox{d}z_b}{z_b}X_b
    F_{g/p}\lr \frac{X_b}{z_b},M_b^2\rr
    \le -\frac{N_C}{2} \tilde{J}_6(z_b)
    + P_{g\leftarrow g}(z_b)\ln\frac{s}{M_b^2} \re \rp
    \nonumber \\
  &&\hspace*{-0.13cm}\lp
    + \int_{X_b}^1 \frac{\mbox{d}z_b}{z_b}X_b
    F_{q/p}\lr \frac{X_b}{z_b},M_b^2
    \rr\le P_{g\leftarrow q}(z_b)\lr\ln\lr\frac{(1-z_b)y_Js}{z_bM_b^2}
    \rr+1\rr\rp\rp \nonumber \\
  &&\hspace*{5cm}\lp\lp -2C_F\frac{1-z_b}{z_b}\re \rg . \nonumber
\eea
The functions $\tilde{J}_i$ are the regular parts of the functions $J_i$
defined in appendix D. They contain the cut-off $y_J$ and are distributions
in $z_b$. The dominant terms for $y_J\rightarrow 0$ are again proportional
to $(\ln^2 y_J)$ and are related to the soft divergence in the parton
initial state. $J_2$ does not contribute after the appropriate
crossing of $(t\leftrightarrow u)$. \\

The results in (70)-(81) are equivalent to those of Baer et al. \cite{xxx7}.
They cannot be compared since these authors introduce a different cut
for integrating over the soft singularities. Instead of invariant masses,
they use a cut on the energy of the gluon. Terms that are independent of
soft singularities agree. For example, this is the case for the
cross section in (79) coming from the collinear photon initial state
singularity.
\section{Inclusive One- and Two-Jet Cross Sections}
In this section, we present some characteristic numerical results for one-
and two-jet cross sections which have been obtained with our method of
slicing the phase space with invariant mass cuts.
In a short communication we have used this method to calculate the
differential two-jet cross section as a function of average rapidity
$\bar{\eta}=\frac{1}{2}(\eta_1+\eta_2)$ of the two jets of highest
transverse energy with the cut on $|\eta_1-\eta_2| < 0.5$ \cite{xxx21}
and compared it to recent experimental data of the ZEUS collaboration
at HERA \cite{xxx22}. For this paper, we have calculated various
one- and two-jet distributions without applying special cuts on
kinematical variables of the initial or final states dictated by the
experimental analysis, although our approach is particularly suitable
for this as has been shown in \cite{xxx21}. \\

The input for our calculations is listed in the following. We have
chosen the CTEQ3M proton structure function \cite{xxx24}, which is a NLO
parametrization with $\overline{\mbox{MS}}$ factorization and
$\Lambda^{(4)} = 239$ MeV. This $\Lambda$ value is also used to calculate
the two-loop $\alpha_s$ value at the scale $\mu = E_T$. The factorization
scale is also $M = E_T$. We do not apply a special cut on the photon energy
fraction $x_a$ in (2) but integrate from $x_{a,\min}$ to $1$.
$Q^2_{\max} = 4~\mbox{GeV}^2$ is the same as used in the jet analysis
of the ZEUS experiment \cite{xxx22}. \\

The further calculation is based on two separate contributions -- a set
of two-body contributions and a set of three-body contributions. Each
set is completely finite, as all singularities have been cancelled
or absorbed into structure functions. Each part depends separately on the
cut-off $y$. If $y$ is chosen large enough, the two parts determine physically
well defined two-jet and three-jet cross sections. Our analytic calculations
are valid only for very small $y$, since terms O$(y)$ have been neglected
in the analytic integrations. For very small $y$, the two pieces have no
physical meaning. In this case, the $(\ln y)$ terms force the two-body
contributions to become negative, whereas the three-body cross sections are
large and positive. In \cite{xxx21}, we have plotted such cross sections
for $y=10^{-3}$. However, when the two- and three-body contributions are
superimposed to yield a suitable inclusive cross section, as for example
the inclusive single-jet cross section, the dependence on the cut-off
$y$ will cancel. Then, the separation of the two contributions with the
cut-off $y$ is only a technical device. The cut-off only serves to
distinguish the phase space regions, where the integrations are done
analytically, from those where they are done numerically. Furthermore,
$y$ must be chosen sufficiently small so that experimental cuts imposed
on kinematical variables do not interfere with the cancellation of the
$y$ dependence. \\

First, we consider the inclusive single-jet cross section. To achieve this,
we must choose a jet definition, which recombines two nearly collinear
partons. As usual, we adopt the jet definition of the Snowmass meeting
\cite{xxx23}. According to this definition, two partons $i$ and $j$ are
recombined if $R_{i,j} < R$, where $R_i = \sqrt{(\eta_i-\eta_J)^2
+(\phi_i-\phi_J)^2}$ and $\eta_J, \phi_J$ are the rapidity and the
azimuthal angle of the recombined jet. We choose $R=1$ in all of the
following results. This means that two partons are considered as two
separate jets or as a single jet depending whether they lie outside or
inside the cone with radius $R$ around the jet momentum. In some cases,
it may occur that two partons $i$ and $j$ qualify both as two individual
jets $i$ and $j$ and as a combined jet $ij$. In this case, we make no
further selection as it is done in \cite{xxx25}, where only the combined
jet is counted. In NLO, the final
state may consist of two or three jets. The three-jet sample consists
of all three-body contributions, which do not fulfill the cone condition. \\

In Fig.~5, the results for the inclusive one-jet cross section
$\mbox{d}^2\sigma/\mbox{d}E_T\mbox{d}\eta$ are shown for various rapidities
$\eta=0, 1$, and $2$ as a function of $E_T$. The NLO and the LO predictions
are plotted. The LO curve is calculated with the same proton structure
function and the same $\alpha_s$ as the NLO curve.
Only the hard scattering cross section
is calculated in LO. For $R=1$, these two cross sections differ only very
little. Of course, the NLO cross section depends on $R$, whereas the LO curve
does not. To separate the curves for the three $\eta$ values, the result
for $\eta=0~(\eta=2)$ has been multiplied by $0.1~(0.5)$. \\

\begin{figure}[htbp]
 \begin{center}
  \begin{picture}(12,8)
   \epsfig{file=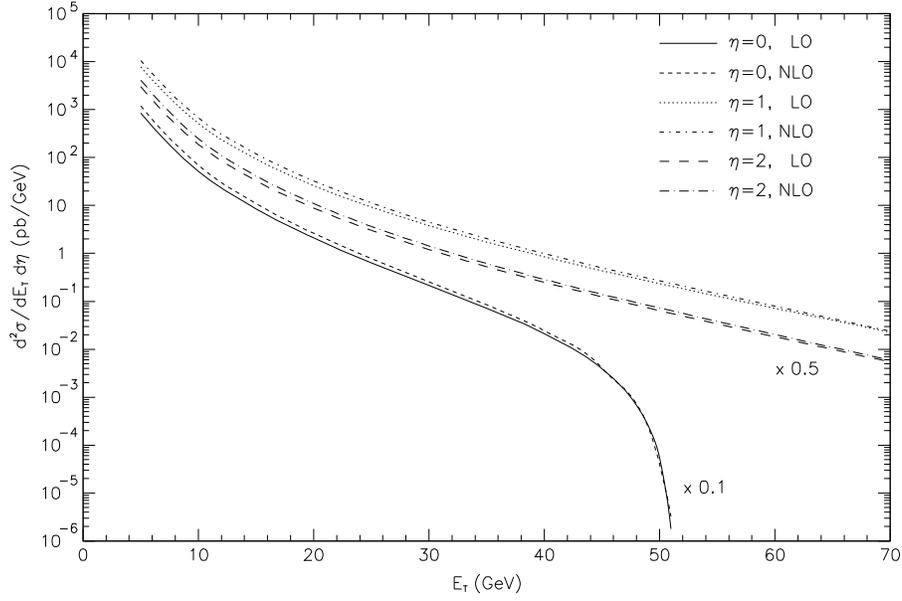,bbllx=102pt,bblly=84pt,bburx=524pt,bbury=712pt,%
           height=8cm,clip=,angle=-90}
  \end{picture}
  \caption{Inclusive single-jet cross section
           $\mbox{d}^2\sigma/\mbox{d}E_T\mbox{d}\eta$
           as a function of $E_T$ for various rapidities $\eta = 0, 1, 2$
           in LO and NLO. The cross section for $\eta = 0~(\eta = 2)$ is
           multiplied by $0.1~(0.5)$.}
 \end{center}
\end{figure}

\begin{figure}[htbp]
 \begin{center}
  \begin{picture}(12,8)
   \epsfig{file=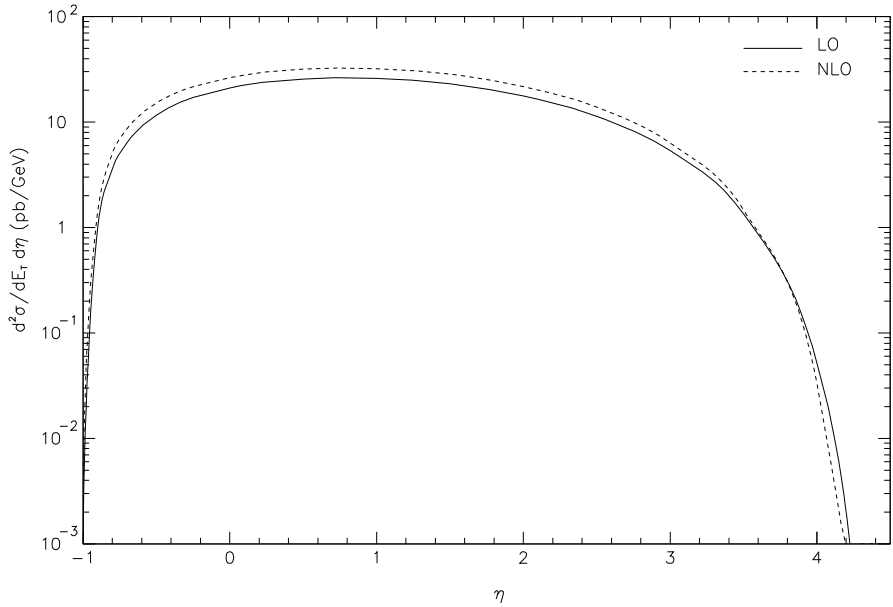,bbllx=102pt,bblly=84pt,bburx=524pt,bbury=712pt,%
           height=8cm,clip=,angle=-90}
  \end{picture}
  \caption{Inclusive single-jet cross section
           $\mbox{d}^2\sigma/\mbox{d}E_T\mbox{d}\eta$
           as a function of $\eta$ for $E_T = 20$ GeV  in LO and NLO.}
 \end{center}
\end{figure}

The maximum of the one-jet cross section is near $\eta=1$. For
$E_T = 20$ GeV, this can be seen from Fig.~6, where
$\mbox{d}^2\sigma/\mbox{d}E_T\mbox{d}\eta$ is plotted as a function
of $\eta$ between $\eta = -1$ and $\eta \simeq 4.2$. At this value
of $E_T$, the kinematical boundaries, where the cross section goes to
zero, differ only little in LO and NLO. The $k$-factor for the NLO
prediction is $k \simeq 1.2$ for the central region of
$0 \leq \eta \leq 3$. \\

Our single-jet distributions have been compared with the earlier results
of B\"odeker \cite{xxx9}, who used the subtraction method to cancel soft
and collinear singularities. We found perfect agreement. To achieve this,
we had to use a rather small value of $y = 10^{-3}$. Furthermore, we had
to add some finite contributions of O$(1)$ for $y \rightarrow 0$, which
correspond to two-jet contributions where an outgoing parton is collinear
to another outgoing or an ingoing parton without the corresponding pole
term in the $2 \rightarrow 3$ matrix elements after partial fractioning.
We also checked that our results are independent of $y$, if $y$ is
sufficiently small. \\

The more essential results are on inclusive two-jet cross sections shown
in the following figures. In Fig.~7, we present
$\mbox{d}^3\sigma/\mbox{d}E_{T_1}\mbox{d}\eta_1\mbox{d}\eta_2$
as a function of $E_{T_1}$ for $\eta_1 = 1$ and various choices of
$\eta_2 = 0, 1, 2$. Here, $E_{T_1}$ and $\eta_1$ are the transverse energy
and the rapidity of the so-called trigger jet. $\eta_2$ is the rapidity
of the second jet, so that $E_{T_1}$ and $E_{T_2}$ are the jets with the
highest transverse momenta for the three-jet contribution. For
exactly two jets in the final state, we have $E_{T_1} = E_{T_2}$.
In Fig.~7, we can see how the cross section decreases when $\eta_2$ is
chosen away from the maximum region at $\eta_2 = 1$. $\eta_1 = 1$ is
always kept fixed. To disentangle the three curves, we have multiplied the
$\eta_2 = 2~(\eta_2 = 0)$ cross section by a factor of $0.5~(0.1)$
as in Fig.~5.
The LO and NLO predictions are very much parallel. The $k$-factor is
$k \simeq 1.2$, similar as in the one-jet cross section. The only large
difference between the LO and NLO cross sections occurs at the boundary
of the phase space. For example, the cross section for $\eta_2 = 0$
has an additional tail in the $E_{T_1}$ distribution coming from
the three-jet cross section, where all three partons lie outside the
cone with radius $R=1$. Such plots can be produced also for other choices
of $\eta_1$ outside the region where the cross section is maximal. \\

\begin{figure}[htbp]
 \begin{center}
  \begin{picture}(12,8)
   \epsfig{file=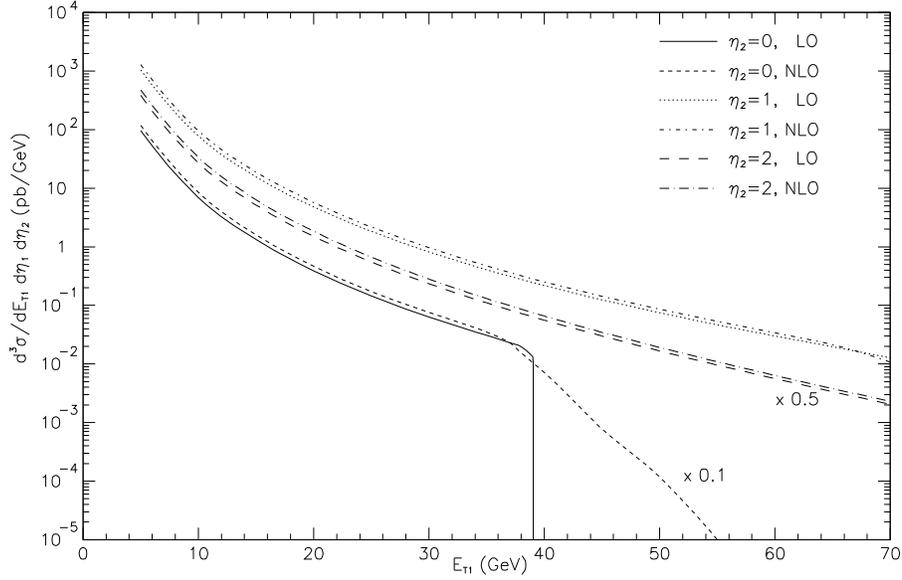,bbllx=102pt,bblly=84pt,bburx=524pt,bbury=712pt,%
           height=8cm,clip=,angle=-90}
  \end{picture}
  \caption{Inclusive dijet cross section
           $\mbox{d}^3\sigma/\mbox{d}E_{T_1}\mbox{d}\eta_1\mbox{d}\eta_2$
           as a function of $E_{T_1}$ for $\eta_1 = 1$ and three values
           of $\eta_2 = 0, 1, 2$. The cross section for $\eta_2=0~(\eta_2
           = 2)$ is multiplied by $0.1~(0.5)$.}
 \end{center}
\end{figure}

We have studied the inclusive two-jet cross section also as a function
of $\eta_1$ and $\eta_2$ for fixed $E_{T_1}$. As an example, we show
in Fig.~8 the two-dimensional distribution
$\mbox{d}^3\sigma/\mbox{d}E_{T_1}\mbox{d}\eta_1\mbox{d}\eta_2$
for $E_{T_1} = 20$ GeV in form of a lego-plot in the intervals
$\eta_1,\eta_2 \in [-1,4]$. Over the whole range of $\eta_1$ and $\eta_2$,
the cross section is approximately symmetric in $\eta_1$ and $\eta_2$.
We show this cross section in LO and NLO, and we see clear differences
between the LO and the NLO prediction. \\

\begin{figure}[htbp]
 \begin{center}
  \begin{picture}(12,8)
   \epsfig{file=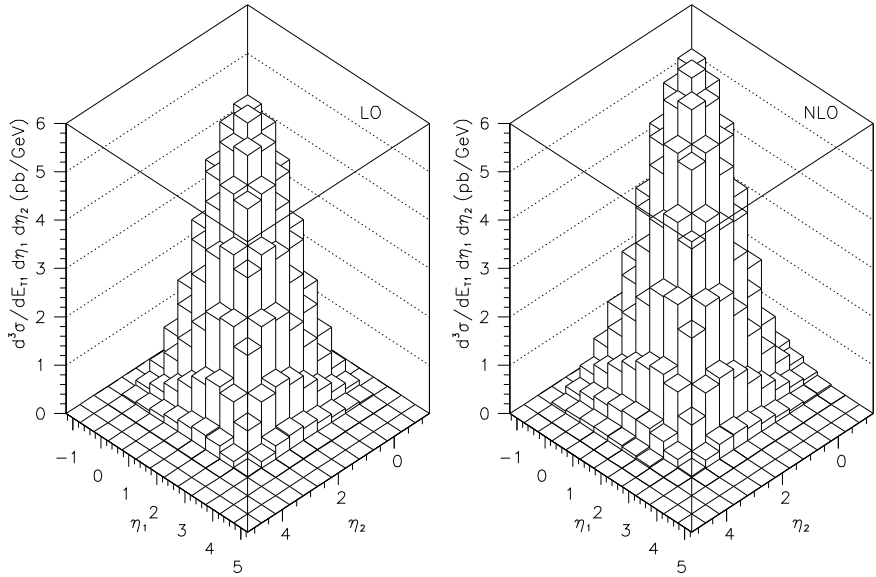,bbllx=102pt,bblly=84pt,bburx=524pt,bbury=712pt,%
           height=8cm,clip=,angle=-90}
  \end{picture}
  \caption{The LO and NLO triple differential inclusive dijet cross section
           for $E_{T_1} = 20$ GeV as a function of $\eta_1$ and $\eta_2$.}
 \end{center}
\end{figure}

This becomes even clearer when
we plot the projections for fixed $\eta_1$ as a function of $\eta_2$.
In Fig.~9, these projections are shown for $\eta_1 = 0, 1, 2$, and $3$
for LO and NLO. We see that in NLO the cross section is always larger than
the LO cross section which is consistent with the $k$-factor larger than
$1$ given above. \\

\begin{figure}[htbp]
 \begin{center}
  \begin{picture}(12,8)
   \epsfig{file=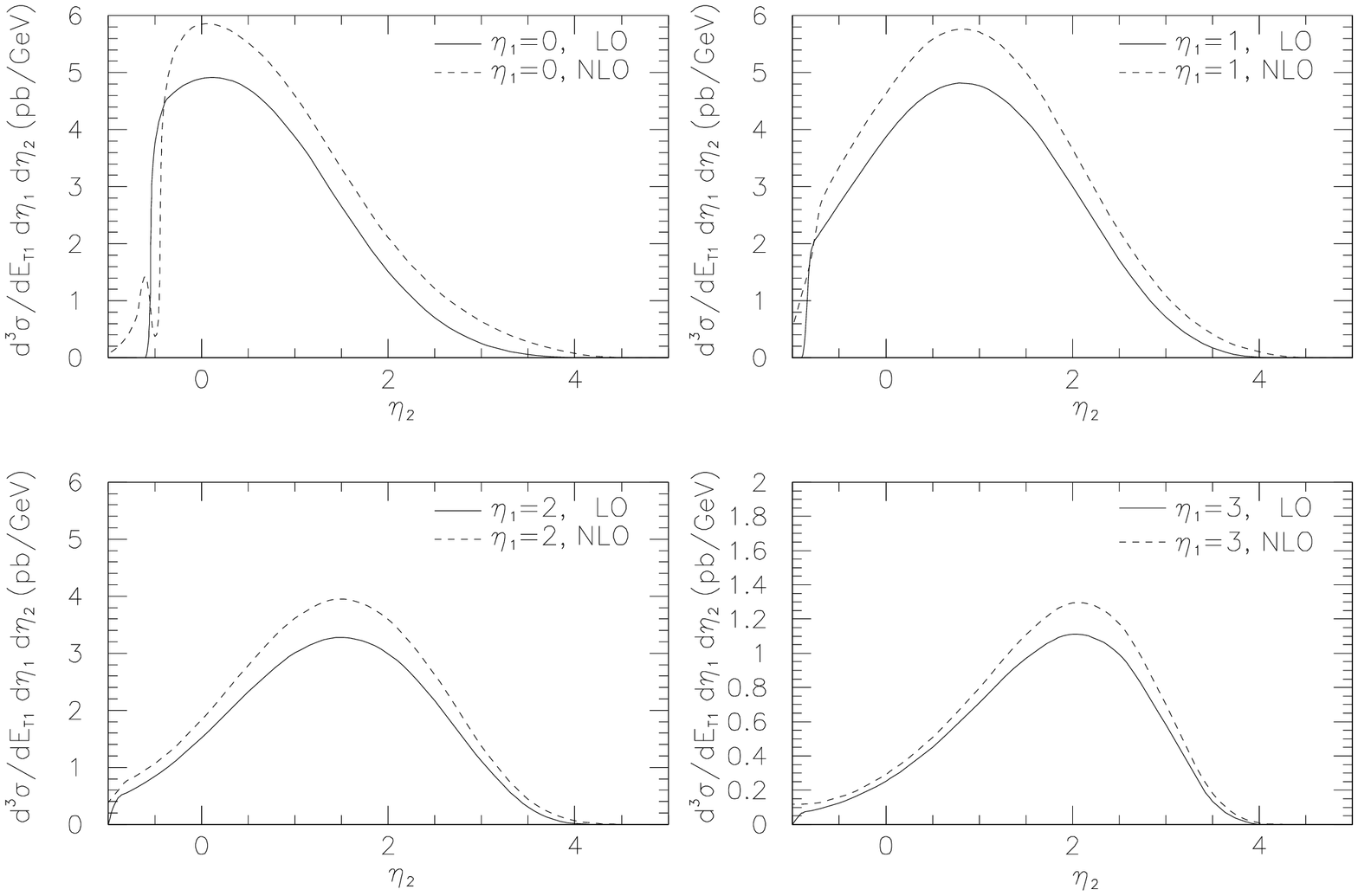,bbllx=102pt,bblly=84pt,bburx=524pt,bbury=712pt,%
           height=8cm,clip=,angle=-90}
  \end{picture}
  \caption{Projections of the LO and NLO triple differential inclusive
           dijet cross section for $E_{T_1} = 20$ GeV as a function
           of $\eta_2$ for fixed $\eta_1 = 0, 1, 2$, and $3$.}
 \end{center}
\end{figure}

It is clear that many more distributions or partially integrated cross
sections using other two-jet variables can be calculated with the
phase space slicing method. As a final application, we have studied the
$R$ and the $y$ dependence of the two-jet cross section, integrated over
$\eta_1, \eta_2$ and $E_{T_1}$ with a minimal $E_{T_1,\min} = 10$ GeV.
The result is shown in Fig.~10, where the LO cross section is independent
of $R$. The NLO cross section is the sum of positive three-body
contributions and negative two-body contributions, both calculated
with the invariant mass cut $y = 10^{-3}$, and this is not independent
of $R$. However, the dependence is rather weak. The cross section increases
with increasing $R$ from $R = 0.1$ to $R = 1$ by a factor of $2$ and
lies above the LO result for $R \geq 0.6$. For $R < 0.1$, the $y$ cut
becomes effective. Therefore we have plotted the curves only for
$R \geq 0.1$. The sum of the two-body and three-body
contributions depends on $R$ like $a + b \ln R + c R^2$ due to
the recombination of two parton momenta $p_i$ and $p_j$ inside the cone
with radius $R$. This way, the singled out transverse energy of a jet can
differ from the individual parton $E_T$'s. This $R$ dependence has been
studied before for the inclusive one-jet cross section with similar
results \cite{xxx8,xxx9}. \\

\begin{figure}[htbp]
 \begin{center}
  \begin{picture}(12,8)
   \epsfig{file=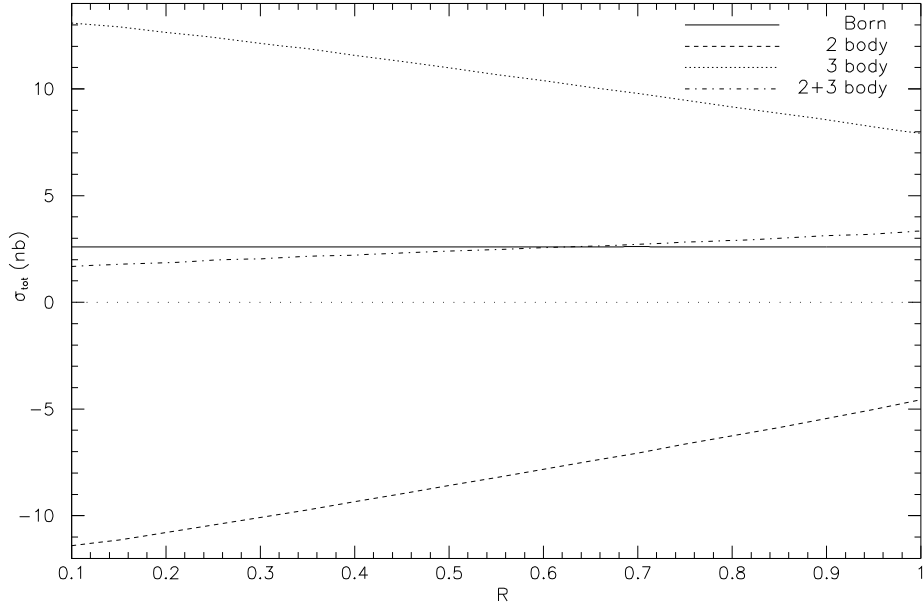,bbllx=110pt,bblly=93pt,bburx=511pt,bbury=706pt,%
           height=8cm,clip=,angle=-90}
  \end{picture}
  \caption{Integrated inclusive dijet cross section for $E_{T_1,\min} = 10$
           GeV as a function of $R$ in NLO (dashed-dotted) and LO (full).
           The 2 body (dashed) and 3 body (dotted) contributions with
           $y = 10^{-3}$ are plotted separately.}
 \end{center}
\end{figure}

We have also calculated the same cross section as a function of the invariant
mass cut $y$ without the cone recombination. In this case, the cut $y$
operates as a physical cut. Of course, the results for the two-body
contributions (being negative for $y < 3\cdot 10^{-2}$) and the three-body
contributions (being above the sum for $y < 3\cdot 10^{-2}$) only make sense
for sufficiently large values of the $y$ cut. For $y \rightarrow 0$, the
two-body (three-body) contributions decrease (increase) like $- \ln^2 y$
($\ln^2 y$). The results are presented in Fig.~11. Here we see that
the sum of both contributions depends only moderately on $y$
($\propto \ln y$) due to the recombination of partons in the final
state. This dependence on $y$ of the summed contribution is equivalent
to the $R$ dependence of the integrated cross section shown in Fig.~10.
We remark that in Fig.~11 the respective cross sections must still be
corrected by finite terms of O($y$). The two-body contribution is
calculated from the analytical formulas given above, where these terms
are neglected. This should influence our results only for $y > 2\cdot
10^{-2}$. It is, however, no problem to calculate these terms numerically. \\

\begin{figure}[htbp]
 \begin{center}
  \begin{picture}(12,8)
   \epsfig{file=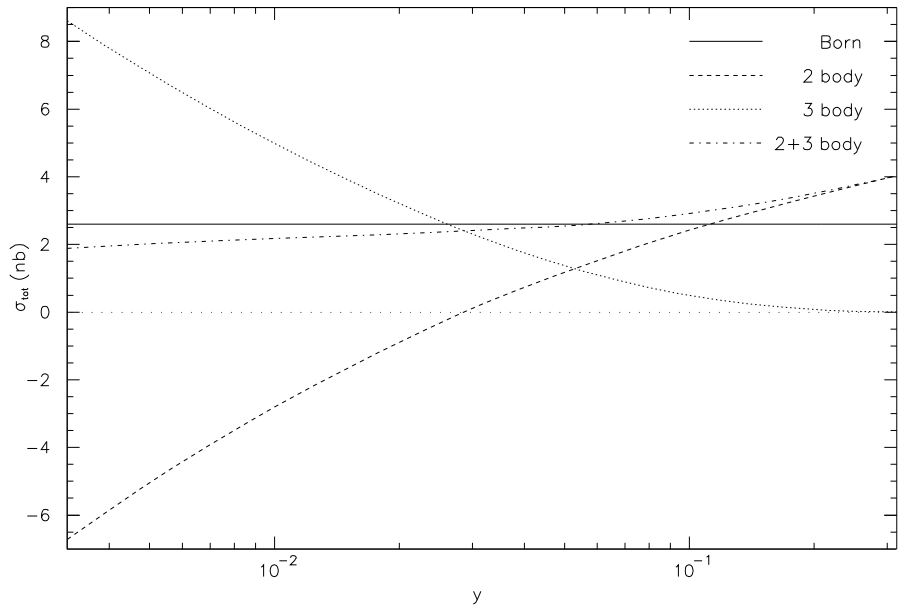,bbllx=109pt,bblly=95pt,bburx=520pt,bbury=706pt,%
           height=8cm,clip=,angle=-90}
  \end{picture}
  \caption{Integrated inclusive dijet cross section for $E_{T_1,\min} = 10$
           GeV as a function of $y$ in NLO (dashed-dotted) and LO (full).
           The 2 body (dashed) and 3 body (dotted) contributions
           are plotted separately.}
 \end{center}
\end{figure}
\clearpage
\section{Summary}
Differential cross sections
$\mbox{d}^3\sigma / \mbox{d}E_T^2\mbox{d}\eta_1\mbox{d}\eta_2$
have been calculated in NLO for direct photoproduction. Infrared and
collinear singularities are cancelled with the phase space slicing method
using an invariant mass cut-off. This method allows to incorporate
various cuts on the final state dictated by the analysis of experimental
data and to perform calculations for different choices of jet
algorithms. Analytical formulas for the different contributions
giving the dependence on the slicing parameter are derived. This is
needed if one wants to develop a complete description of
photoproduction events including hadronization and other corrections.
Numerical results for the two-jet inclusive cross sections
at HERA have been presented, and LO and NLO predictions have been compared.
For a cone radius of $R = 1$, the NLO corrections lead to an increase
of the order of 20\% compared to the LO prediction which is
calculated with the same proton structure function and the same
$\alpha_s$ as the NLO prediction.
\setcounter{equation}{0}
\clearpage
\begin{appendix}
\renewcommand{\theequation}{\mbox{\Alph{section}.\arabic{equation}}}
\section{Virtual Corrections}
In this appendix, we give the interference terms of the virtual corrections
with the LO QCD Compton scattering $\gamma q \rightarrow gq~(=T_q(s,t,u))$
and the LO photon-gluon fusion $\gamma g \rightarrow q\bar{q}~
(=T_g(s,t,u))$ matrix elements as defined in equations (16) and (17).
The results depend only on the Mandelstam variables $s$, $t$, and $u$:
\bea
  V_{q1} & = &
    \le -\frac{2}{\epsilon^2} +\frac{1}{\epsilon}
    \lr 2\ln \ts - 3 \rr +\frac{2\pi^2}{3} -7 +\ln^2\tu \re T_q(s,t,u)
  \\ &&
    {}-2\ln\us+4\ln\ts-3\frac{s}{u}\ln\us-\lr 2+\frac{u}{s} \rr
    \lr \pi^2+\ln^2\tu \rr-\lr 2+\frac{s}{u} \rr \ln^2\ts ,
  \nonumber \\ &&
  \nonumber \\
  V_{q2} & = &
    \le \frac{2}{\epsilon^2}+\frac{1}{\epsilon}
    \lr 2\ln\ts-2\ln\us \rr
      +\frac{\pi^2}{3}+\ln^2\tu \re T_q(s,t,u)
   \\ &&
    {}-2\ln\us+4\ln\ts-\lr 2+\frac{u}{s} \rr
    \lr \pi^2+\ln^2\tu \rr -\lr 2+\frac{s}{u} \rr \ln^2\ts ,
  \nonumber \\ &&
  \nonumber \\
  V_{g1} & = &
    \le -\frac{2}{\epsilon^2}-\frac{3}{\epsilon}
    +\frac{2\pi^2}{3}-7+\ln^2\ts+\ln^2\us \re T_g(s,t,u)
   \\ &&
    {}+2\ln\ts+2\ln\us+3\frac{u}{t}\ln\ts+3\frac{t}{u}\ln\us
    +\lr 2+\frac{u}{t} \rr \ln^2\us+\lr 2+\frac{t}{u}\rr \ln^2\ts ,
  \nonumber \\ &&
  \nonumber \\
  V_{g2} & = &
    \le \frac{2}{\epsilon^2}+\frac{1}{\epsilon}
    \lr -2\ln\ts-2\ln\us \rr
      +\frac{\pi^2}{3}+\ln^2\frac{tu}{s^2} \re T_g(s,t,u)
   \\ &&
    {}+2\ln\ts+2\ln\us +\lr 2+\frac{u}{t}\rr \ln^2\us +\lr 2+\frac{t}{u} \rr
    \ln^2\ts .
   \nonumber
\eea
\setcounter{equation}{0}
\section{Final State Corrections}
This appendix contains the real corrections coming from the integration
of the $2 \rightarrow 3$ matrix elements over the final state singular
region of phase space. In addition to the Mandelstam variables,
the result depends on the integration cut-off $y_F$:
\bea
  F_{1} & = & \le \frac{1}{\epsilon^2}+\frac{1}{\epsilon}
            \left( -\ln\ts + \frac{3}{2} \right)
            -\frac{1}{2}\ln^2\ts+2\ln y_F\ln\ts
            -2\mbox{Li}_2\lr\frac{ y_F s}{t}\rr -\frac{\pi^2}{3}
            +\frac{7}{2} \rp \\
      &   & \lp {}-\ln^2 y_F-\frac{3}{2}
            \ln y_F \re C_F^2 T_q(s,t,u)
            , \nonumber \\
      &   & \nonumber \\
  F_{2} & = & \le -\frac{1}{\epsilon} \ln\ts -\frac{1}{2}\ln^2\ts
            +2\ln y_F\ln\ts
            -2\mbox{Li}_2\left( \frac{ y_F s}
            {t} \right) \re \lr -\frac{1}{2}N_CC_F \rr T_q(s,t,u), \\
      &   & \nonumber \\
  F_{3} & = & \le -\frac{2}{\epsilon^2}+\frac{1}{\epsilon}
            \left( \ln\us -2 \right) -4+2\ln y_F+\ln^2 y_F+\ln^2
            \frac{ y_F s}{-u} -\frac{1}{2}\ln^2\us
            + \frac{2\pi^2}{3}\rp \\
      &   & \lp {}+ 2\mbox{Li}_2\left( \frac{ y_F s}
            {u} \right) \re \lr -\frac{1}{2}N_CC_F \rr
            T_q(s,t,u), \nonumber \\
      &   & \nonumber \\
  F_{4} & = & \le -\frac{5}{3\epsilon} +\frac{5}{3}\ln y_F-
            \frac{31}{9} \re \lr -\frac{1}{2}N_CC_F \rr T_q(s,t,u) , \\
      &   & \nonumber \\
  F_{5} & = & \le -\frac{1}{3\epsilon} -\frac{5}{9}+
            \frac{1}{3}\ln y_F \re C_F T_q(s,t,u) , \\
      &   & \nonumber \\
  F_{6} & = & \le \frac{2}{\epsilon^2}+\frac{3}{\epsilon}
            -\frac{2\pi^2}{3}+7-2\ln^2 y_F
            -3\ln y_F \re \frac{C_F}{2} T_g(s,t,u), \\
      &   & \nonumber \\
  F_{7} & = & \le \frac{1}{\epsilon}\ln\frac{tu}{s^2}
            +\ln^2\frac{ y_F s}{-t} +\ln^2\frac{ y_F s}{-u}
            -2\ln^2 y_F-\frac{1}{2}\ln^2\ts-\frac{1}{2}\ln^2\us
            +2\mbox{Li}_2\lr\frac{y_Fs}{t}\rr \rp \\
      &   & \lp{}+2\mbox{Li}_2\lr\frac{y_Fs}{u}\rr
            \re \lr -\frac{N_C}{4} \rr T_g(s,t,u). \nonumber
\eea
\setcounter{equation}{0}
\section{Photon Initial State Corrections}
Like gluons, real photons can split into $q\bar{q}$ pairs. After integration
over the collinear region of phase space, the Born matrix elements for
$q\bar{q}\rightarrow gg$ ($=T_{1a,b}(s,t,u))$, $qq'\rightarrow qq'$
($=T_2(s,t,u))$, and $qq\rightarrow qq$ ($=T_{2,3}(s,t,u))$
can be factorized out. The result depends on the Mandelstam variables,
on the integration cut-off $y_I$, and on the integration variable
$z_a$:
\bea
  I_{1}&=&\le-\frac{1}{\epsilon}\frac{1}{2N_C}P_{q_i\leftarrow \gamma}(z_a)
           +\frac{1}{2N_C}P_{q_i\leftarrow \gamma}(z_a)
           \ln\lr\frac{1-z_a}{z_a}y_I\rr+\frac{Q_i^2}{2}
           \re 2C_F^2T_{1a}(s,t,u), \\
  I_{2}&=&\le-\frac{1}{\epsilon}\frac{1}{2N_C}P_{q_i\leftarrow \gamma}(z_a)
           +\frac{1}{2N_C}P_{q_i\leftarrow \gamma}(z_a)
           \ln\lr\frac{1-z_a}{z_a}y_I\rr+\frac{Q_i^2}{2}
           \re (-N_CC_F)T_{1b}(s,t,u), \\
  I_{3}&=&\le-\frac{1}{\epsilon}\frac{1}{2N_C}P_{q_i\leftarrow \gamma}(z_a)
           +\frac{1}{2N_C}P_{q_i\leftarrow \gamma}(z_a)
           \ln\lr\frac{1-z_a}{z_a}y_I\rr+\frac{Q_i^2}{2}
           \re 2C_FT_2(s,t,u), \\
  I_{4}&=&\le-\frac{1}{\epsilon}\frac{1}{2N_C}P_{q_i\leftarrow \gamma}(z_a)
           +\frac{1}{2N_C}P_{q_i\leftarrow \gamma}(z_a)
           \ln\lr\frac{1-z_a}{z_a}y_I\rr+\frac{Q_i^2}{2}
           \re 2C_FT_2(t,s,u), \\
  I_{5}&=&\le-\frac{1}{\epsilon}\frac{1}{2N_C}P_{q_i\leftarrow \gamma}(z_a)
           +\frac{1}{2N_C}P_{q_i\leftarrow \gamma}(z_a)
           \ln\lr\frac{1-z_a}{z_a}y_I\rr+\frac{Q_i^2}{2}
           \re 2\frac{C_F}{N_C}T_{3}(s,t,u) \\
       &&  {}+\mbox{zycl. permutations of s,t, and u}, \nonumber \\
  I_{6}&=&\le-\frac{1}{\epsilon}\frac{1}{2N_C}P_{q_i\leftarrow \gamma}(z_a)
           +\frac{1}{2N_C}P_{q_i\leftarrow \gamma}(z_a)
           \ln\lr\frac{1-z_a}{z_a}y_I\rr+\frac{Q_i^2}{2}
           \re (-2C_F)T_{1a}(t,s,u) \\
       &&  {}+(t\leftrightarrow u), \nonumber \\
  I_{7}&=&\le-\frac{1}{\epsilon}\frac{1}{2N_C}P_{q_i\leftarrow \gamma}(z_a)
           +\frac{1}{2N_C}P_{q_i\leftarrow \gamma}(z_a)
           \ln\lr\frac{1-z_a}{z_a}y_I\rr+\frac{Q_i^2}{2}
           \re N_CT_{1b}(t,s,u) \\
       &&  {}+(t\leftrightarrow u). \nonumber
\eea
The Altarelli-Parisi splitting function $P_{q_i\leftarrow \gamma}$ is equal
to $2N_CQ_i^2P_{q\leftarrow g}$ (see Appendix D).
\setcounter{equation}{0}
\section{Parton Initial State Corrections}
In this appendix, the parton initial state corrections are given as
a function of the Mandelstam variables $s$, $t$, $u$, of the cut-off
$y_J$, and of the additional integration variable $z_b$:
\bea
  J_{1}&=& \le -\frac{1}{\epsilon}\frac{1}{C_F} P_{q\leftarrow q}(z_b)
            +\delta (1-z_b) \lr \frac{1}{\epsilon^2}
            +\frac{1}{\epsilon} \lr -\ln \ts +\frac{3}{2} \rr
            +\frac{1}{2}\ln^2\ts+\pi^2\rr\right. \\
        & & {}+1-z_b+(1-z_b) \ln \lr \frac{1-z_b}{z_b}y_J\rr
            +2R_+\lr\ts\rr-2\ln\lr\ts\lr\frac{1-z_b}{z_b}\rr ^2\rr
            \nonumber \\
        & & \lp {}-2\frac{z_b}{1-z_b}\ln\lr 1+\frac{-t}{y_Js}
            \frac{1-z_b}{z_b}\rr
            \re C_F^2T_q(s,t,u) ,
            \nonumber \\
        & & \nonumber \\
  J_{2}&=& \le \delta (1-z_b) \lr \frac{1}{\epsilon} \ln \ut
            +\frac{1}{2}\ln^2\ts-\frac{1}{2}\ln^2\us\rr
            +2R_+\lr\ts\rr -2R_+\lr\us\rr\rp\\
        & & {}-2\ln\lr\ts\lr\frac{1-z_b}{z_b}\rr ^2\rr
            +2\ln\lr\us\lr\frac{1-z_b}{z_b}\rr ^2\rr\nonumber \\
        & & \lp {}-2\frac{z_b}{1-z_b}\ln\lr 1+\frac{-t}{y_Js}
            \frac{1-z_b}{z_b}\rr
            +2\frac{z_b}{1-z_b}\ln\lr 1+\frac{-u}{y_Js}
            \frac{1-z_b}{z_b}\rr
            \re \nonumber \\
        & & \lr -\frac{1}{2}N_CC_F\rr T_q(s,t,u), \nonumber \\
        & & \nonumber \\
  J_{3}&=& \le-\frac{1}{\epsilon}\frac{1}{C_F}P_{g\leftarrow q}(z_b)
            +\frac{1}{C_F}P_{g\leftarrow q}(z_b)\ln\lr\frac{1-z_b}{z_b}y_J\rr
            -2\frac{1-z_b}{z_b} \re \frac{C_F}{2}T_g(s,t,u) , \\
        & & \nonumber \\
  J_{4}&=& \le -\frac{2}{\epsilon} P_{q\leftarrow g}(z_b)
            +2P_{q\leftarrow g}(z_b)\ln\lr\frac{1-z_b}{z_b}y_J\rr+1
            \re C_FT_q(s,t,u) , \\
      &   & \nonumber \\
  J_{5}&=& \mbox{O}(\epsilon) , \\
      &   & \nonumber \\
  J_{6}&=& \le \frac{2}{\epsilon}\frac{1}{N_C}P_{g\leftarrow g}(z_b)
            +\delta(1-z_b)
            \lr -\frac{2}{\epsilon^2}+\frac{1}{\epsilon}
            \lr \ln\frac{tu}{s^2}-\frac{2}{N_C}
            \lr\frac{11}{6}N_C-\frac{1}{3}N_f\rr\rr\rp\rp \\
        & & \lp{}-\frac{1}{2}\ln^2\ts-\frac{1}{2}\ln^2\us-2\pi^2\rr
            -2R_+\lr\ts\rr-2R_+\lr\us\rr \nonumber \\
        & & {}+2\ln\lr\ts\lr\frac{1-z_b}{z_b}\rr^2\rr
            +2\ln\lr\us\lr\frac{1-z_b}{z_b}\rr^2\rr\nonumber \\
        & & {}+2\frac{z_b}{1-z_b}\ln\lr 1+\frac{-t}{y_Js}
            \frac{1-z_b}{z_b}\rr
            +2\frac{z_b}{1-z_b}\ln\lr 1+\frac{-u}{y_Js}
            \frac{1-z_b}{z_b}\rr\nonumber \\
        & & \lp{}-4\lr\frac{1-z_b}{z_b}+z_b(1-z_b)\rr
            \ln\lr\frac{1-z_b}{z_b}y_J\rr
            \re \lr -\frac{N_C}{4}\rr T_g(s,t,u) , \nonumber
\eea
where we have introduced
\bea
R_+(x) & = & \lr \frac{\ln \lr x \lr
             \frac{1-z_b}{z_b}\right)^2\right)}{1-z_b}\right)_+
\eea
for convenience. The Altarelli-Parisi splitting functions are defined
as
\bea
  P_{q\leftarrow q} (z_b) & = &
    C_F \left[ \frac{1+z_b^2}{(1-z_b)_+} + \frac{3}{2} \delta (1-z_b)
    \right] , \\
  P_{g\leftarrow q} (z_b) & = &
    C_F \left[ \frac{1+(1-z_b)^2}{z_b} \right] , \\
  P_{g\leftarrow g} (z_b) & = &
    2 N_C \left[ \frac{1}{(1-z_b)_+}+\frac{1}{z_b}+z_b(1-z_b)-2 \re
    + \left[ \frac{11}{6}N_C-\frac{1}{3}N_f\right] \delta (1-z_b),\\
  P_{q\leftarrow g} (z_b) & = &
    \frac{1}{2} \left[ z_b^2+(1-z_b)^2 \right]
. \eea
As the integration over $z_b$ runs from $X_b$ to $1$, the $+$ distributions
in this paper are defined as
\beq
  D_+[g] = \int_{X_b}^1 \mbox{d}z_b D(z_b) g(z_b)
          -\int_0^1     \mbox{d}z_b D(z_b) g(1)
, \eeq
where
\beq
  g(z_b) = \frac{1}{z_b} F_{b'/p}\lr\frac{X_b}{z_b}\rr h(z_b),
\eeq
and $h(z_b)$ is a regular function of $z_b$ \cite{xxx20}.
This leads to additional
terms not given here explicitly when (D.12) is transformed so that both
integrals are calculated in the range $[X_b,1]$.
\setcounter{equation}{0}
\end{appendix}
\newpage

\end{document}